\documentclass[acmtog,natbib]{acmart}
\usepackage{booktabs} 

\usepackage[ruled]{algorithm2e} 

\SetAlFnt{\small}
\SetAlCapFnt{\small}
\SetAlCapNameFnt{\small}
\SetAlCapHSkip{0pt}
\IncMargin{-\parindent}

\def\E{\mathbb{E}}

\def\mat#1{{\left(\begin{array}{cccccccccccccccccccccccccccc}#1\end{array}\right)}}

\setcopyright{rightsretained}

\begin{document}

\title{Geometric Computing with Chain Complexes  
\subtitle{\em Design and Features of a Julia Package} }

\author{Francesco Furiani}
\affiliation{%
  {Dept. of Mathematics and Physics,}
  {Roma Tre University,}
  {Rome,}
  {Italy}
}
\author{Giulio Martella}
\affiliation{%
  {Dept. of Engineering,}
  {Roma Tre University,}
  {Rome,}
  {Italy}
}
\author{Alberto Paoluzzi}
\affiliation{%
  {Dept. of Mathematics and Physics,}
  {Roma Tre University,}
  {Rome,}
  {Italy}
}


\begin{abstract}
Geometric computing with chain complexes allows for the computation of the whole chain of linear spaces and (co)boundary operators generated by a space decomposition into a cell complex. The space decomposition is stored and handled with LAR (Linear Algebraic Representation), i.e. with sparse integer arrays, and allows for using  cells of a very general type, even non convex and with internal holes. In this paper we discuss the features and the merits of this approach, and describe the goals and the implementation of a software package aiming at providing for simple and efficient computational support of geometric computing with any kind of meshes, using linear algebra tools with sparse matrices. The library is being written in Julia, the novel efficient and parallel language for scientific computing. This software, that is being ported on hybrid architectures (CPU+GPU) of last generation, is yet under development.
\end{abstract}

%
%


%
%


\thanks{This work is partially supported  from SOGEI S.p.A. ---  the ICT company of the Italian Ministry of Economy and Finance, by grant 2016-17, and by the  ERASMUS+ EU project medtrain3dmodsim. 

  Authors addresses: F.~Furiani and A.~Paoluzzi, Department of Mathematics and Physics, Roma Tre University; 
  G.~Martella, Department of Engineering, Roma Tre University.
}

\renewcommand{\shortauthors}{Geometric Computing with Chain Complexes}

\nonstopmode

\maketitle

\section{Introduction}
\subsection{What, When, Why}

In the four decades-long history of solid modeling, providing computer representations for mathematical models of manufactured objects, the standard characterization of a solid normally required some specialized data structure, called a Brep~\cite{Hoffmann}, to represent a cell decomposition of the boundary. When non-manifold solids are considered, in order to make their operations closed with respect to object configurations with touching boundaries, the computer representation becomes much more convoluted, as having to state and maintain the topological consistence  between  incidence relations of boundary cells, and the cyclic orderings between adjacent pairs. 

In this paper we discuss a largely different approach. Our computer model of a solid (a) may contain a decomposition of either the boundary or the interior of the solid; (b) the type of cells of the decomposition (i.e.~their shape or topology) is not fixed a priori; so allowing for (c) a direct association of semantics to the cell; and providing (d) a very general covering of most application domains that require a computer model of the geometric shape and physical behaviour of the object. This approach encompasses (i) computer graphics and rendering, (ii) solid modeling and its common operations, (iii) mesh decompositions and physical simulations, as well as (iv)  imaging for 3D medical and other applications. Even more, the computer representation itself does not need some weird data structures, typical of non-manifold descriptions, but only requires few very sparse binary matrices (either one or two, depending on the topology of cells), used to codify the cells as subsets of vertices, i.e., to store the subset of vertices on the boundary of each cell.

The software package introduced in this paper is based on homological algebra of chain complexes. We can find conceptual roots in the description of physical world behaviour through geometric shape and its properties, and in the reduction of differential geometry and physical laws to topological facts~\cite{Whitney:1957,Tonti:1976}. We believe that a discrete  topological  framework may support all/most physical simulations~\cite{Tonti:2013,Tonti:2014:WSD:2563304.2563867,Desbrun03discreteexterior},  simply by leveraging the coboundary operator for all conservation / balance / compatibility / equilibrium type laws,  and linear (metric) operators for constitutive / measured laws~\cite{vadim:2017}.  

The development of a  literate software~\cite{doi:10.1093/comjnl/27.2.97} experimental library supporting LAR, the Linear Algebraic Representation~\cite{DBLP:journals/cad/DiCarloPS14} of topology, started with Python in 2012, and was interrupted in 2016 by A.P.'s lack of time. Few months later we started a new development project in Julia, the novel efficient computer language for scientific computing~\cite{BEKS14}, with F.F.~and G.M.~as main developers. The package core, computing the bases and the transformations of the exact sequence of linear spaces (chain complex)  induced by a collection of cellular complexes, merged to generate a single arrangement of the Euclidean space, will be discussed here. The porting of other modules of the package from Python to Julia started in this semester at Roma Tre University.

\subsection{Chain-based computing}

Chain-based modeling and computing is based on representation of $d$-cell subsets as \emph{chains}, elements of linear spaces $C_d$.  Their dual spaces of cochains are defined as linear combinations of maps from elementary chains (cells) to $\mathbb{R}$.
Chains and cochains are represented as sparse arrays, \emph{ergo} as dense arrays of indices and/or values, the simplest and more efficient data structure of languages oriented to scientific computing.
Therefore, basic operations on chains as vectors (sum and product times a scalar) are implemented over  sparse/dense arrays.

Often we must deal with \emph{oriented} cells,  naturally represented as  sparse/dense arrays of \emph{signed} indices/values. This representation applies to very general kinds of connected cells: $p$-simplicies, $p$-cubes, polytopal, non-convex, even punctured, i.e., homeomorphic to cells with internal holes. 
Given an ordering over bases of $p$-cells, (linear) topological operators, like boundary and coboundary (see Section~\ref{background}), and hence discrete gradients, curls and divergence, are represented as integer or real matrices.
Incidence operators between chain spaces of different dimension are easily computed by matrix products of characteristic matrices (see Section~\ref{matrices}), possibly transposed.

Data validity is straightforward to test by checking for satisfaction of basic equations $\partial\partial=\emptyset$ of a chain complex.
Since both characteristic and operator matrices are very sparse, their products are computed with  specialized algorithms for sparse matrices, whose complexity is roughly linear in the size of the output sparse matrix, i.e., in the number of its stored non-zero elements.
Furthermore, several software packages are available for efficient linear algebra with sparse matrices, and most of such softwares are already ported to last-generation mixed architectures (CPU + GPU).

Because of the above points (sparse-array- and GPU-based), the chain-complex approach seems to fit  with Neural Networks architectures, where typical data types for input/output from neurons are dense or sparse arrays. 
Novel operators are being investigated in this framework; e.g., the topological traversal of a set of boundary cycles, returning the chain of interior cells.
This operator is a sort of pseudoinverse of the boundary operator, which goes from chains to boundary cycles.

It is worthwhile noting that incidence queries and other types of geometric/topological computations are not performed element-wise, that necessarily require iterative or recursive programming patterns, but just by matrix multiplication times chains (sets of cells, i.e., sparse arrays), so adapting naturally to parallel and/or dataflow computational patterns found in HPC and CNN architectures. 

Last but not least, a chain-based representation crosses the boundaries of many  geometric computing subfields. It would be easy to show that chains of cells may represent \emph{any subset} of 2D/3D images, as well as both boundary and decompositive schemes for solid modeling, as well as data structures used for domain meshing and physical simulation, as wells as the data structures used for geographical systems and outdoor/indoor mapping, and so on. That is not surprising, since chain-complex-based computing of geometric data may take for granted more than one century of algebraic topology methods.

\subsection{Previous Work}

An up-to-date review of the mapping schemes used to provide computer representations of manufactured objects can be found in the dedicated survey chapter on Solid Modeling, by Hoffmann and Shapiro, in the Handbook of Discrete and Computational Geometry~\cite{Goodman:2017:HDC:285869}. The challenges introduced by the novel frontiers of computational modeling of material structures are discussed in~\cite{DBLP:journals/cad/RegliRSS16}.
A description of the computational geometry CGAL software~\cite{Fabri:2000:DCC:358668.358687}, implementing 2D/3D arrangements with Nef polyhedra~\cite{Hachenberger:2007:BOS:1247750.1248141} and the construction of arrangements of lines, segments, planes and other geometrical objects can be found in~\cite{fhktww-a-07}.  
{Discrete Exterior Calculus (DEC) with simplicial complexes was introduced by \cite{Hirani:2003:DEC:959640,Desbrun03discreteexterior} and made popular by \cite{Desbrun:2006:DDF:1185657.1185665,Elcott:2006:BYO:1185657.1185666}.}
More recently,  a systematic recipe   has been proposed  in~\cite{Zhou:2016:MAS:2897824.2925901} for constructing a family of exact constructive solid geometry operations starting from a collection of triangle meshes. Conversely, the dimension-independent approach, data structures and algorithms implemented in \texttt{LARLIB.jl} and discussed here, may apply to any collection of both boundary and decompositive representations, including 2D/3D computer images and solid meshes.

\subsection{Paper Contents}

In Section~\ref{background} we shortly recall the main concepts regarding LAR (Linear Algebraic Representation) as a general representation scheme for geometric and topological modeling, and the main features of Julia language, including its native support for parallel processing.
In Section~\ref{larlib} after describing a list of subpackages of the Python prototype library, we discuss the design goals of the new Julia implementation, and present the \emph{Merge} algorithm.
In Section~\ref{examples} we show by examples the meaning of both the characteristic matrix of a basis of $d$-cells, and the (co)boundary matrix, with some examples of 2D and 3D chain complex computation.
In the Conclusion section the next development steps are outlined, and our view of the evolution of this software is provided.

\section{Background}\label{background}

A large amount of research was invested in the last few years into the rewriting 
of standard algorithms over very-large graphs, i.e., cellular
\emph{1-complexes} including 0-cells (nodes) and 1-cells (edges), by using linear algebra methods based
on sparse matrices~\cite{Buono:2016:OSM:2925426.2926278,doi:10.1177/1094342011403516,Kepner:2011:GAL:2039367,Davis:2006:DMS:1196434}. Our approach allows to work with linear algebra and sparse matrices over cellular \emph{$d$-complexes}.

\subsection{Characteristic matrices and (co)boundary operators}\label{matrices}

Let $X$ be a topological space, and
$\Lambda(X) = \bigcup_p\Lambda_p$ ($p \in {0, 1,\ldots,d}$) be a
{cellular decomposition} of $X$, with $\Lambda_p$ a set of {closed} and connected $p$-cells. 
A \emph{CW-structure} on the space $X$ is a filtration
$\emptyset = X_{-1} \subset X_0 \subset X_1 \subset \ldots \subset X = \bigcup_p X_p$,
such that, for each $p$, the \emph{skeleton} $X_p$ is homeomorphic
to a space obtained from $X_{p-1}$ by attachment of $p$-cells in
$\Lambda_p(X)$. A \emph{cellular complex} is a space $X$ provided with a CW-structure.

With abuse of language, we consider a finite cellular complex $X$ as generated by a discrete partition of an Euclidean space. In computing a cellular complex as the space \emph{arrangement} of a collection of geometric objects $\mathcal{S}$, i.e.~when  $X := \mathcal{A}(\mathcal{S})$, we actually compute the whole \emph{chain complex} $C_\bullet$ generated by $X$, i.e.:

\begin{figure}[htbp] 
   \centering
   \includegraphics[width=0.45\linewidth]{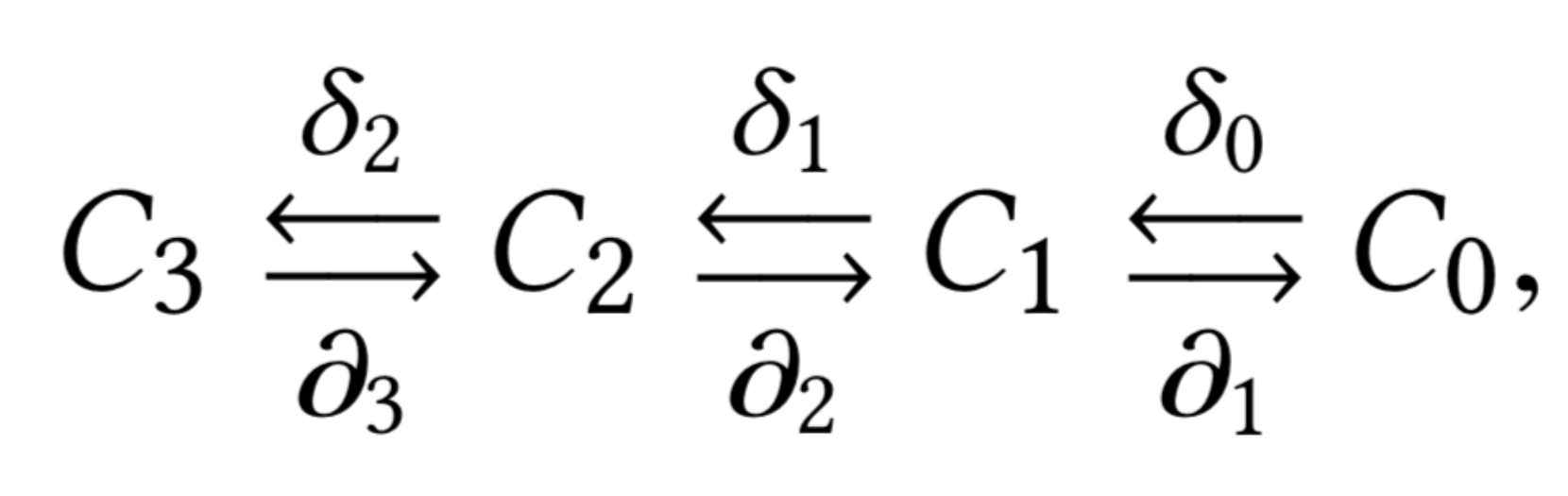} 
   \label{fig:chains}
\end{figure}
\vspace{-4mm}
\noindent
where $C_p$ ($0\geq p\geq 3$) is a linear space of \emph{$p$-chains} (subsets  of $p$-cells with algebraic structure).  The linear operators $\partial_p$ and $\delta_p$ are called \emph{boundary} and \emph{coboundary}, respectively, with 
\[
\partial_{p-1}\circ\partial_p  = \emptyset = \delta_{p}\circ\delta_{p-1} ,
\]
and where 
\[
\delta_{p-1} = \partial_p^\top, \qquad 1\leq p\leq 3.
\] 
For a discussion about the identification, performed here, between chains and cochains, see~\cite{2017arXiv170400142P}.

A \emph{characteristic function} $\chi_A: S\to\{0,1\}$ is a function defined on a finite set $S=\{s_j\}$, that indicates membership of an element $s_j$ in a subset $A\subseteq S$, having the value 1 for all elements of $A$ and the value 0 for all elements of $S$ not in $A$. 
We call \emph{characteristic matrix} $M$ of a collection of subsets $A_i\subseteq S$ ($i=1,\ldots,n$) the binary  matrix $M=(m_{ij})$, with $m_{ij} = \chi_{A_i}(s_j)$, that provides a basis for a linear space $C_p$ of \emph{$p$-chains}. 

\subsection{Sparse matrices as representations}

Our  \emph{representation scheme}~\cite{Requicha:1980:RRS:356827.356833}, i.e.~our mapping between mathematical models of solids and their computer representations, uses linear \emph{chain spaces} $C_p$ as models, and sparse \emph{characteristic matrices} $M_p$  of $p$-cells as symbolic representations, where the $p$-cell $\sigma^k\in\Lambda_p$ is represented as the $k$-th binary row of the  sparse characteristic matrix $M_p: C_0\to C_p$. Since $M_p$ matrices are \emph{very} sparse, we can compactly represent the basis of a $C_p$ space, when the ordering of $0$-cells (vertices) and $p$-cells have been fixed, as an \emph{array}, indexed by cell indices, \emph{of arrays} of indices of vertices. Vertex positions are represented by a 2-array of points with $d$ real coordinates. The reader may find in~\ref{sec:appendix} such representation of both the input and the output cellular complexes of Figure~\ref{fig:appendix}. 

It may be interesting to note that our mathematical calculation of the space arrangement (see Figure~\ref{fig:rubik-2}) generated by the  \emph{Merge} of a collection of complexes~\cite{2017arXiv170400142P} returns the whole chain complex $C_\bullet$, i.e., both the bases of chain spaces $C_p$ as (sparse) characteristic matrices, and their connecting (co)boundary operators $\partial_p$ as sparse matrices.

\subsection{Features of LAR (Linear Algebraic Representation)}\label{lar}

In development since several years, in a joint collaboration between 
Roma Tre University and the University of Wisconsin
at Madison, the chain-based modeling approach~\cite{DiCarlo:2009:DPU:1629255.1629273,ieee-tase} 
provides a complete representation
of solid models taking advantage of linear homological algebra, and using only sparse matrices.

The LAR scheme~\cite{Dicarlo:2014:TNL:2543138.2543294} provides the capability of using few pivotal algorithms---basically sparse matrix-vector multiplication~\cite{gemmexp}---to compute and analyse the space arrangements and chain complexes generated by cellular complexes. In
particular, this approach produces linear operators to compute 
incidences between cells of any \(p\) dimension (\(0\leq p\leq d\)), allowing for fast
algebraic computation of boundary, coboundary and mutual incidence of {any}
\emph{subsets of cells}, called \emph{chains}, to consider as elements of linear spaces of
chains~\cite{hatcher:2002}.

Also, LAR can be used for both boundary representation and cellular
decomposition of objects, so unifying the management of engineering
meshes, 2D/3D images, solid and graphical models in both 2D and 3D, and
even in higher dimensions.
In particular, this scheme can be used for simplicial, cubical and polytopal
(e.g.~Voronoi) meshes, and for cellular decompositions with any shape of
cells, also including holes of any dimension. In a fair comparison with
\emph{Brep} solid models~\cite{Woo:85,Baumgart:1972:WEP:891970}, 
LAR needs (before the possible bitwise compression) a
space proportional to twice the number of boundary edges. Similar
improvements hold for decompositive modeling schemes.

\subsection{Scientific Computing with Julia Language}

\emph{Julia}~\cite{BEKS14} is a novel language for scientific computing, aiming at producing high performance code based on readable and compact sources, in a programming style similar to Python or Matlab. Julia also provides an excellent support for parallel and distributed computing environments using both low-level and high-level accelerated libraries. 
In few words, Julia is fast, easy to write and to read, and ready for direct support of several kinds of scientific computations with its continuously growing ecosystem of dedicated libraries.
The \texttt{LARLIB.jl} package aims to become the Julia's library for dealing with big geometric environments and models, and with meshes of any type.

\section{Larlib Package}\label{larlib}

The \texttt{LARLIB.jl} project aims are (a) to provide a parallel implementation
of the novel  technique~\cite{2017arXiv170400142P} to compute the \(d\)-cells of the
unknown space arrangement generated by a collection of ($d$--1)-dimensional geometric
structures, (b) to supply some frequently used numerical and topological computing tools to other geometric/topological algorithms, and (c) to apply this technology and software to the 
multilevel extraction of structures and models emerging from
high-resolution biomedical imaging~\cite{doi:10.1080/16864360.2016.1168216}, in the framework of next generation
neural networks and image analysis research.

The breadth of the LAR scheme~\cite{Dicarlo:2014:TNL:2543138.2543294}, planned by design to replace with sparse matrices and
linear algebra the intricate data structures and algorithms used for
non-manifold solid modeling, has a broad range. The application
opportunities opened by this approach seem pretty large, spanning from games to 
product lifecycle management (PLM), from biochemical to biomedical applications, and from  3D modeling  and printing  from images to scene understanding.

\subsection{Python Prototype}

A first Python prototype was written using literate programming methods~\cite{doi:10.1093/comjnl/27.2.97}, to make the code more easily accessible to other developers and/or  more easily extended or ported. In particular, we used Python 2.7, \LaTeX\ and Nuweb. The  package and its documentation are accessible at the url~\texttt{https://github.com/cvdlab/lar-cc}.
 \texttt{Larlib} requires  \texttt{Scipy}, \texttt{Pyplasm}, \texttt{PyOpenGL}, and the \texttt{Triangle} packages, and includes several modules, described in the following.

\begin{description}
\item[boundary]
To convert between between dimension-in\-dep\-end\-ent, dimension-dependent, oriented and non ori\-ent\-ed operators;

\item[integr]
Signed and non-signed finite integration of polynomials on simplicial and/or cellular 2-chains and 3-chains, using the divergence theorem for volume integration;

\item[lar2psm]
Back- and forth-conversion from/to hierarchical polyhedral complex (HPC) data structure (\texttt{pyplasm} class) and LAR (\texttt{numpy} sparse arrays). It also provides access to \texttt{pyplasm} interactive visualizer of big geometric data sets, and to cell decorators with indices, used for testing;

\item[larcc]
Basic data structures and conversions between several types of sparse array representation;

\item[largrid]
Implementation of Cartesian product of general cellular complexes, used also for multidimensional extrusion of complexes, 
generation of multidimensional cuboidal grids and their skeletons, and for parametric mapping (charts) of manifolds;

\item[larstruct]
Definition of structured spaces as hierarchical assemblies of complexes and affine transformations, and traversal algorithms,  with a semantics derived from  PHIGS, standard ISO for 3D graphics;

\item[mapper]
Generation of 1D, 2D, 3D manifold cellular complexes from parametric coordinate functions mapped on simplicial decompositions of chart domains;

\item[morph]
\emph{Mathematical Morphology} over 2D and 3D images represented as cellular complexes, including main operators of erosion, dilation, opening and closing, implemented with algebraic operations on chains as image subsets;

\item[simplexn]
Combinatorial generation of well-defined simplicial complexes and grids, and extraction of their facets and  $p$-dimensional skeletons ($0\leq p\leq d$). Includes combinatorial simplicial extrusion, with complexity $\Omega(n)$, where $n$ is the output size;

\item[splines]
Tensor product (and other) methods for generation of polynomial and rational (curve, surface, and solid) splines. Includes generation of profile products of curves, and generation of generalized cones and cylinders.
\end{description}


\begin{figure*}[htbp] 
   \centering
   \includegraphics[width=0.2\linewidth]{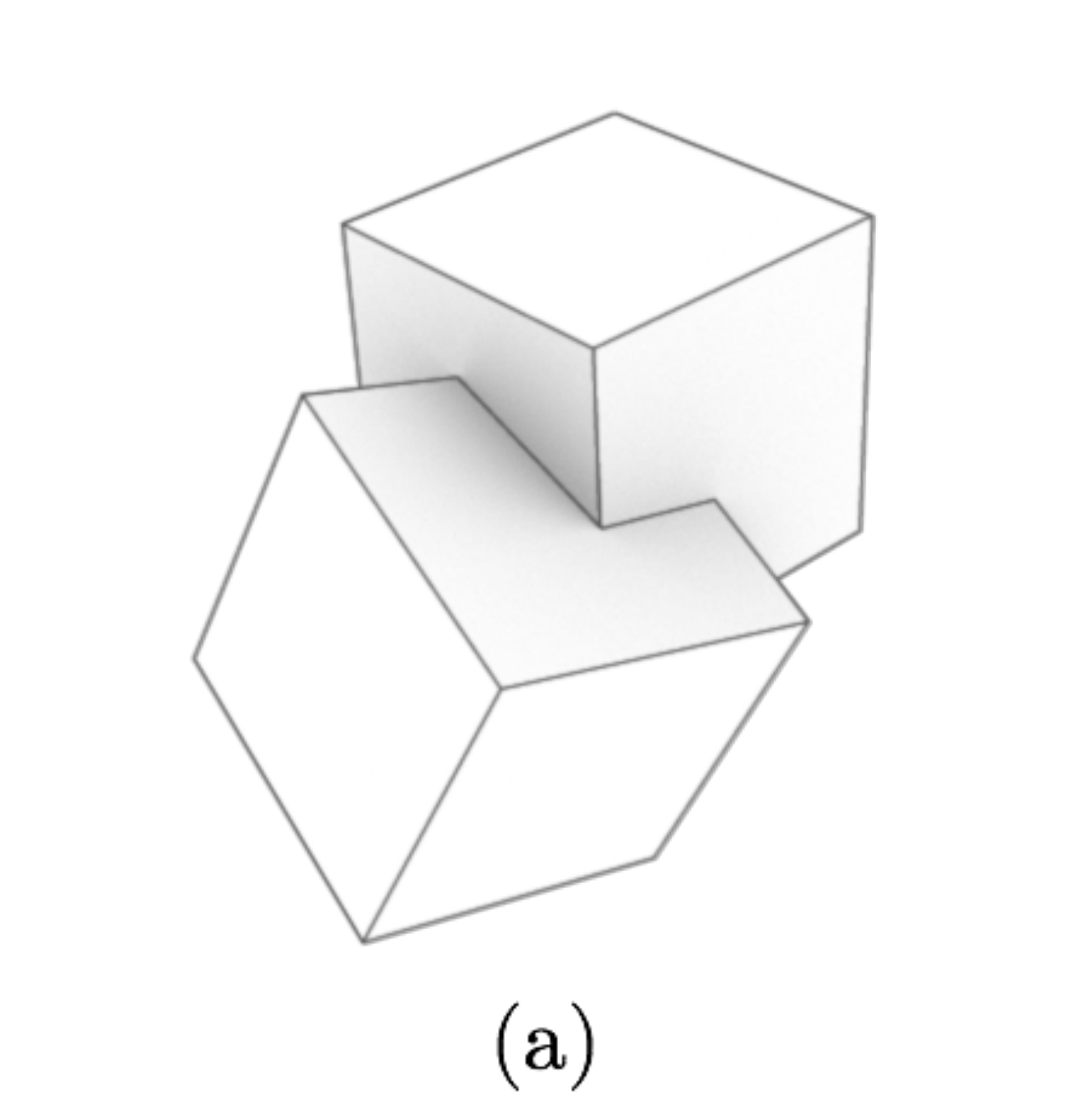}%
   \includegraphics[width=0.2\linewidth]{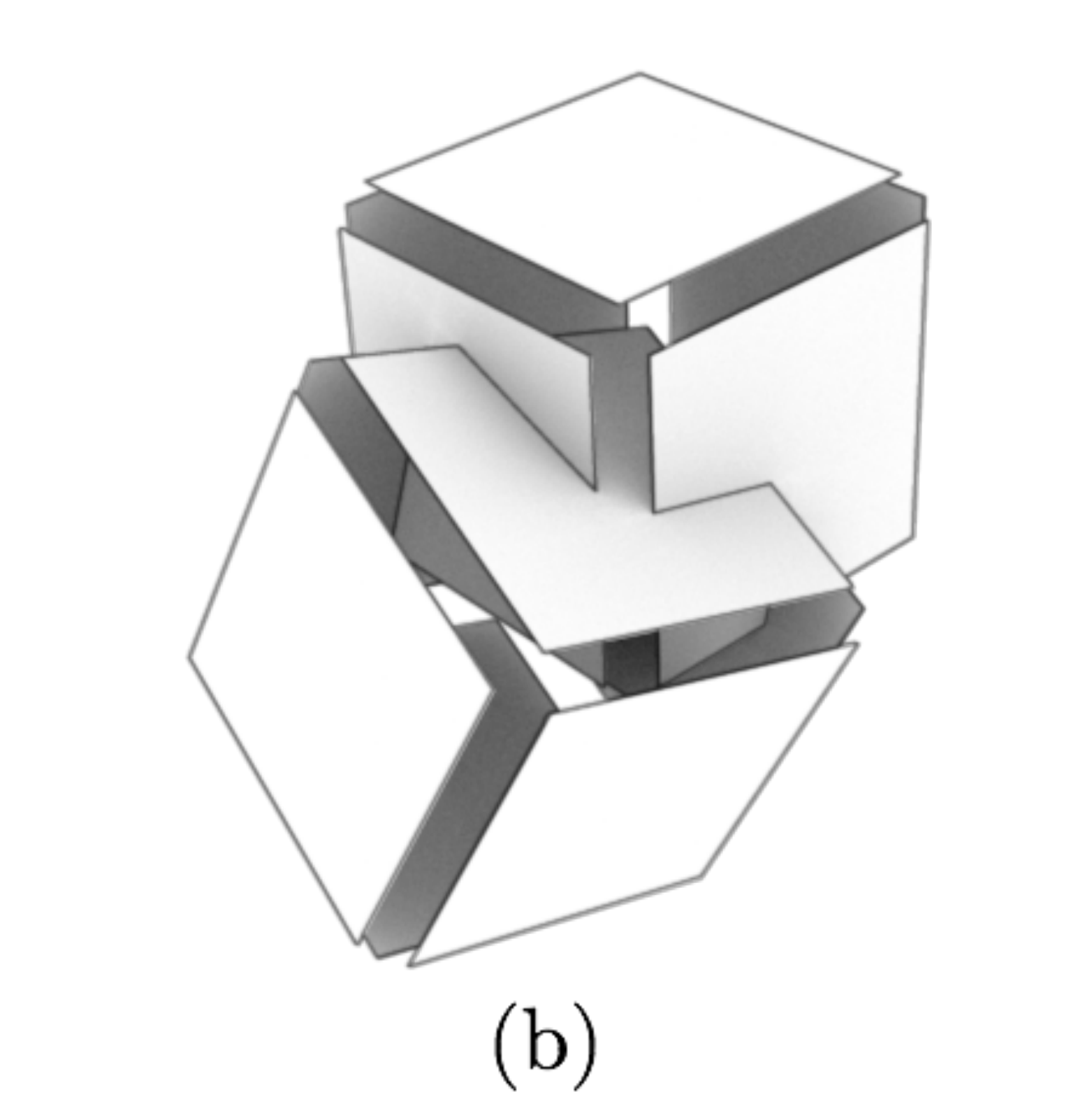}%
   \includegraphics[width=0.2\linewidth]{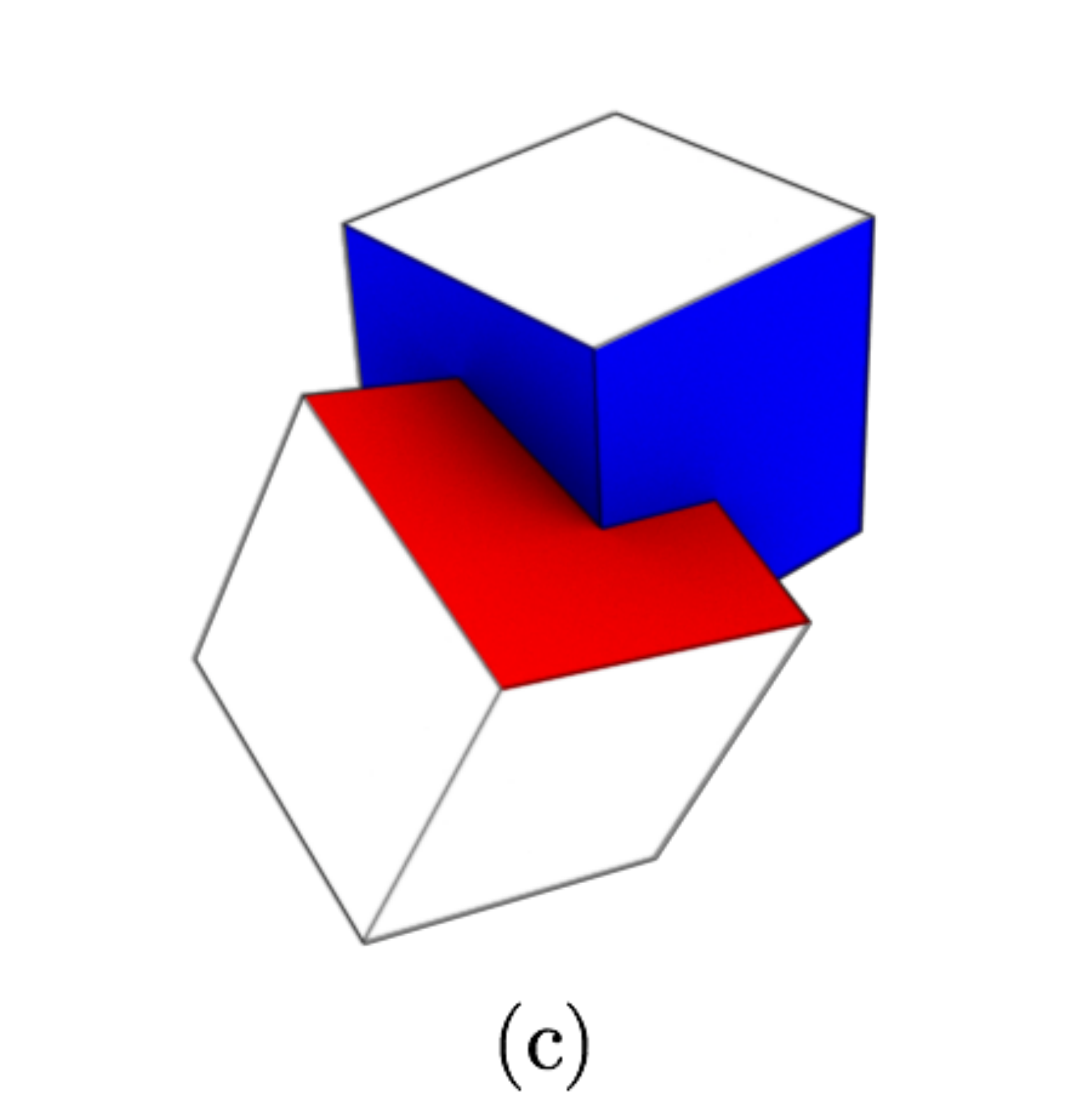}%
   \includegraphics[width=0.2\linewidth]{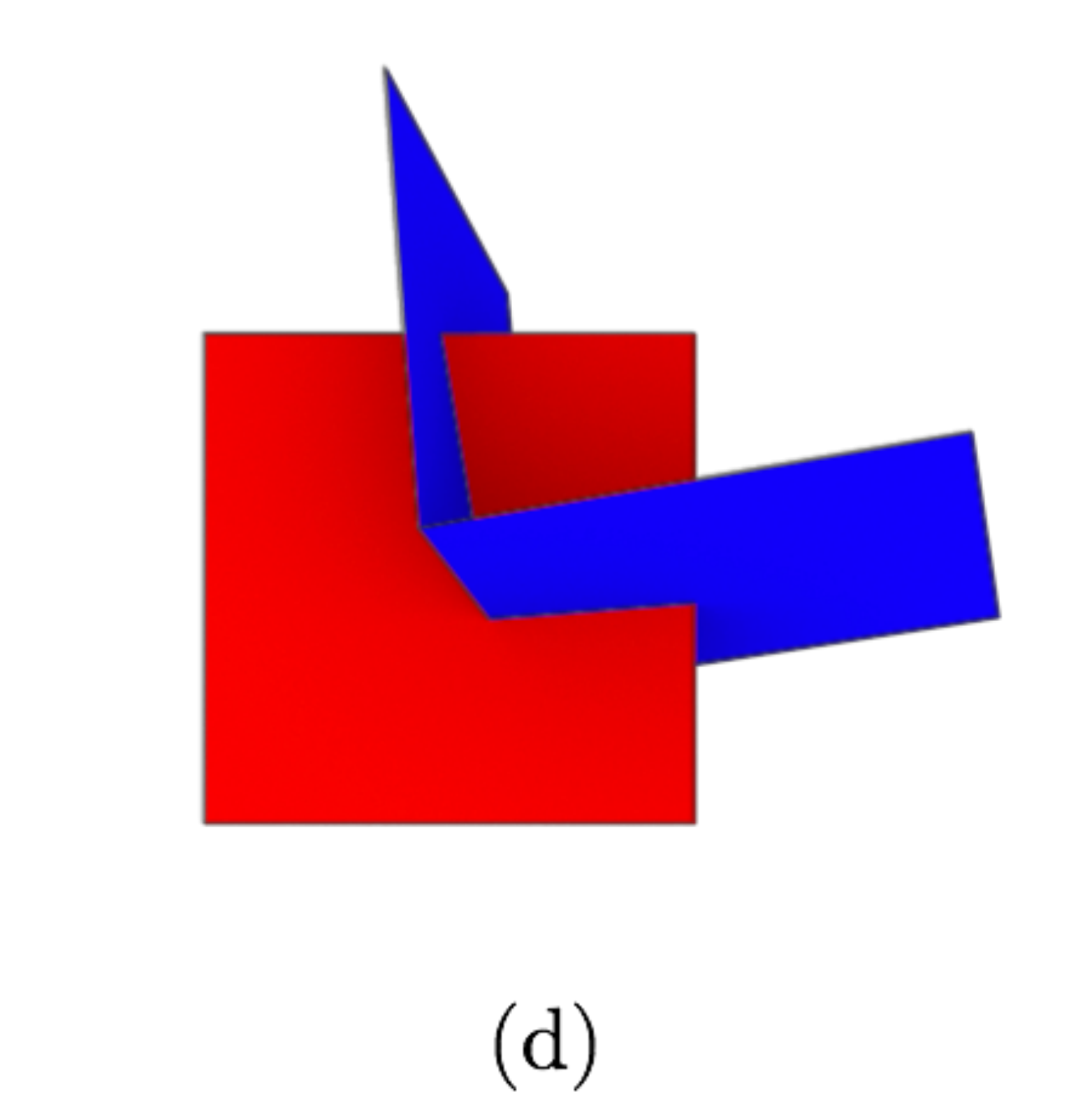}%
   
   \includegraphics[width=0.2\linewidth]{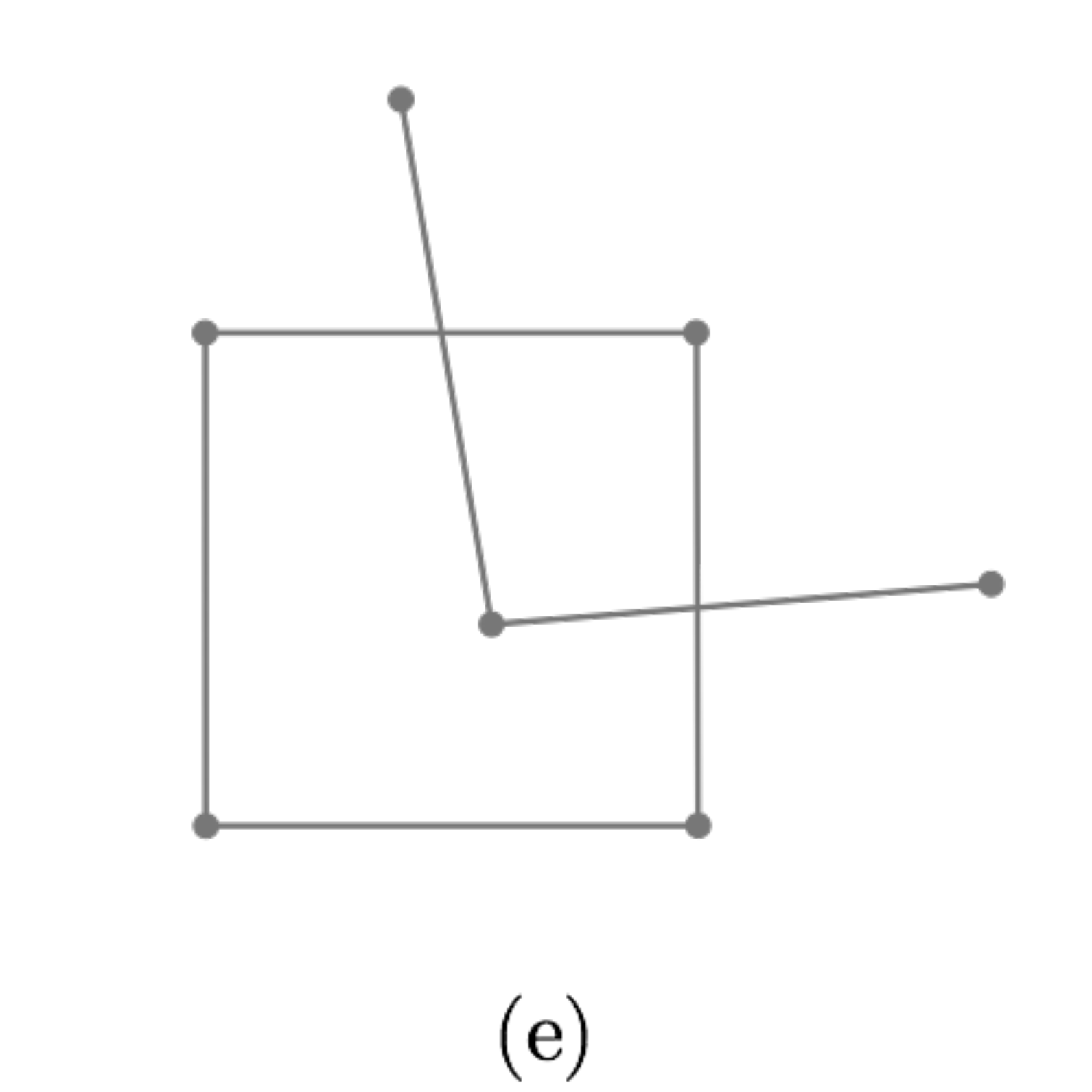}%
   \includegraphics[width=0.2\linewidth]{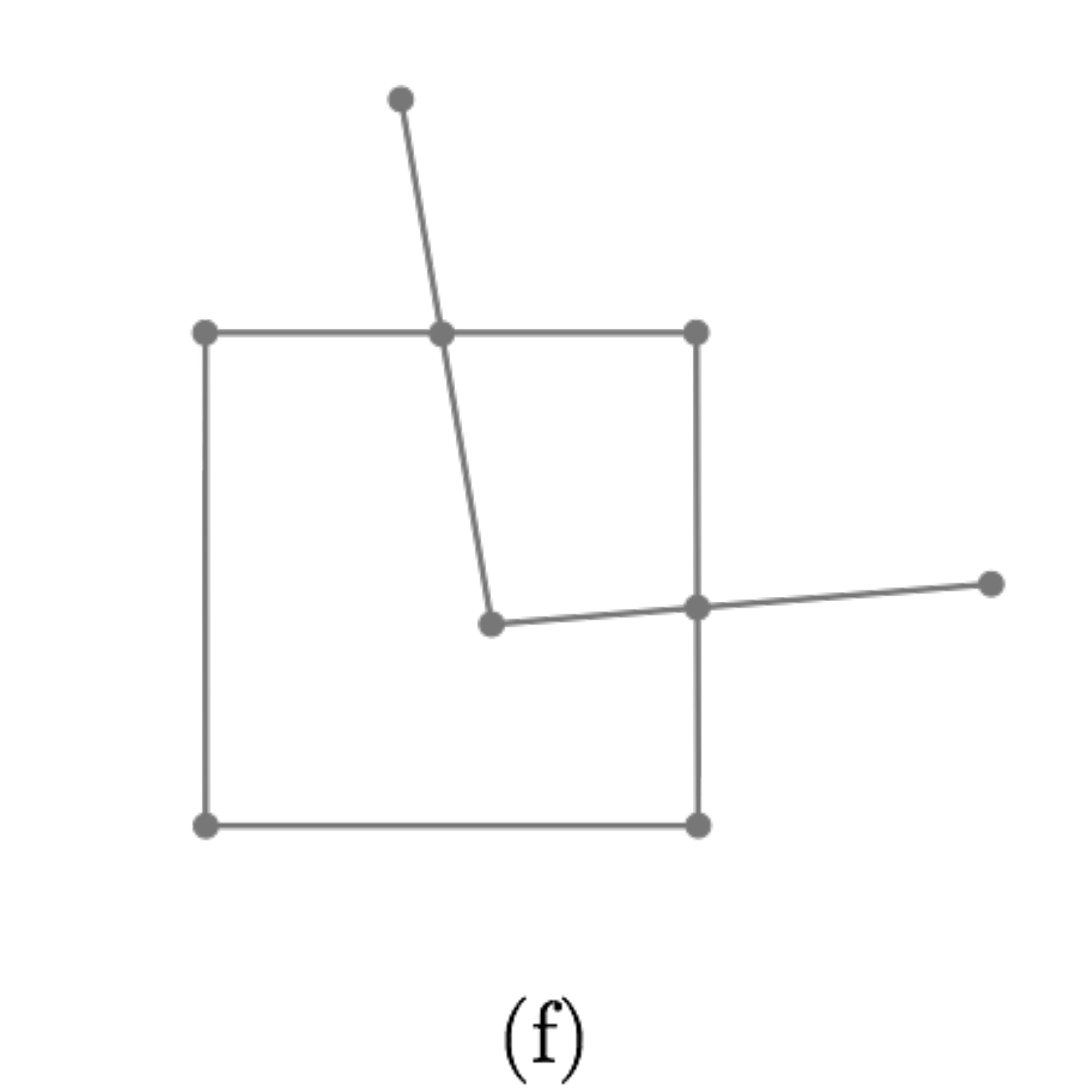}%
   \includegraphics[width=0.2\linewidth]{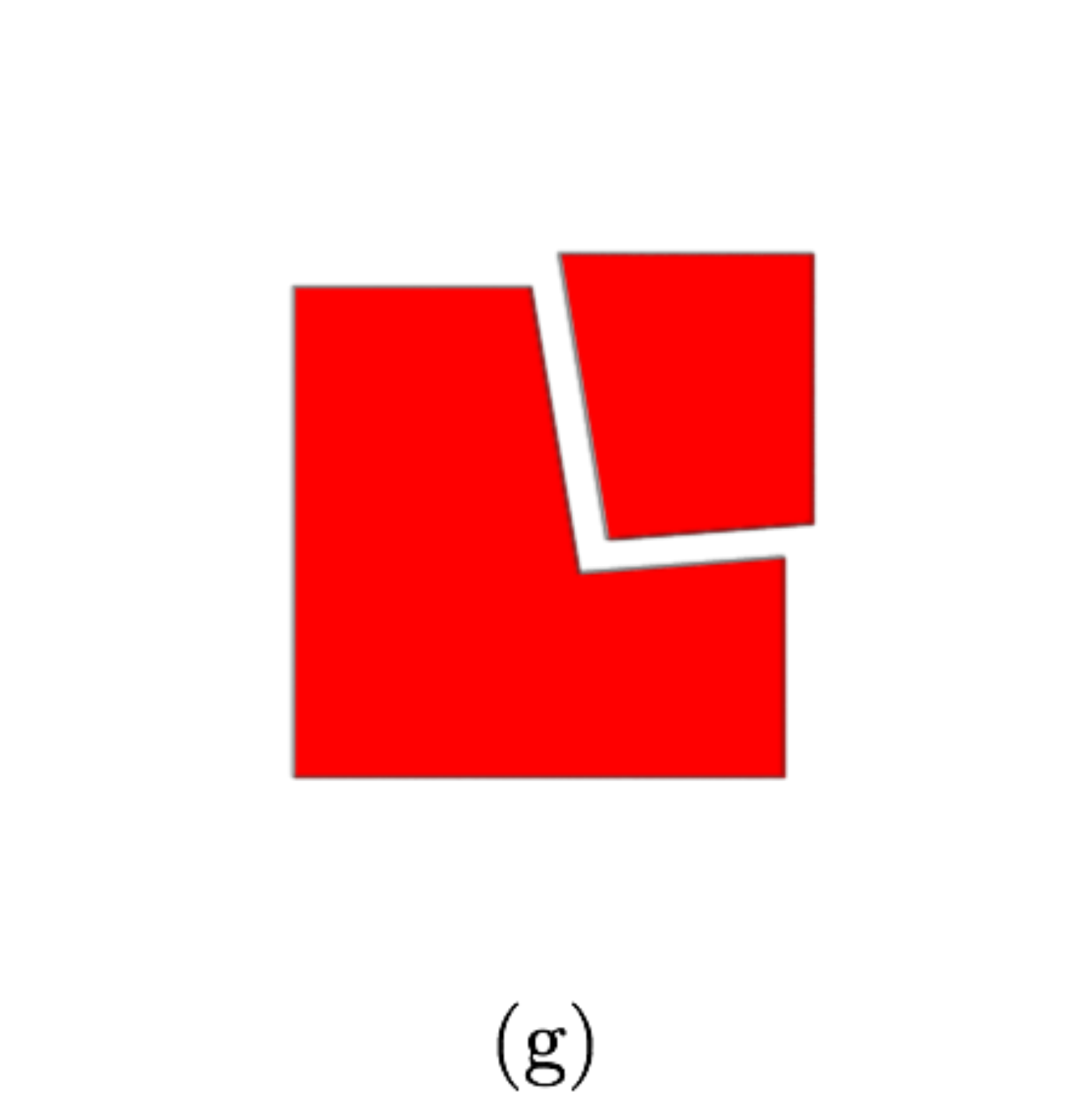}%
   \includegraphics[width=0.2\linewidth]{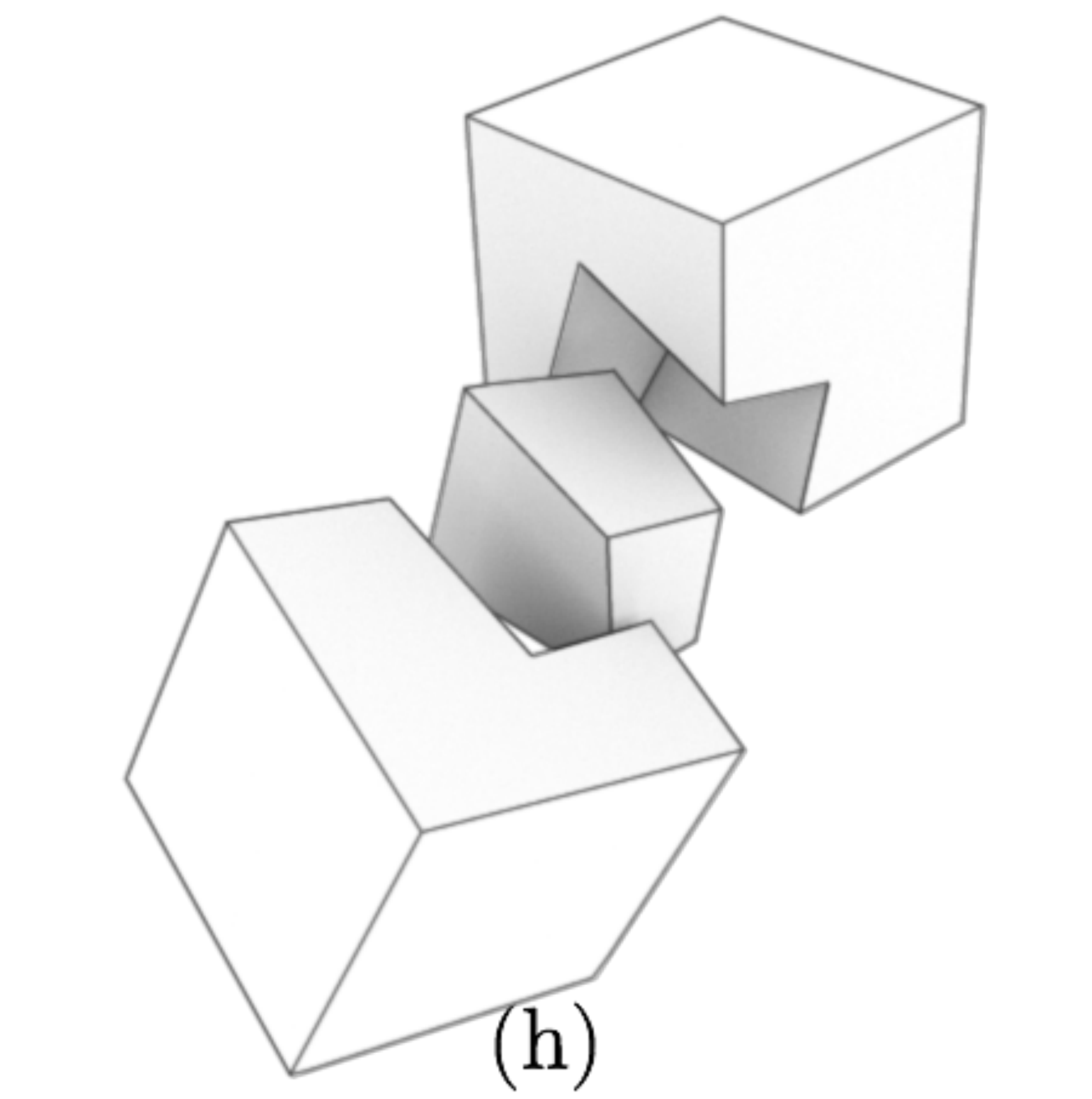}%
   
   \caption{Cartoon display of the Merge algorithm: (a) the two input solids; (b) the exploded input collection $B$ of 2-cells embedded in $\E^3$; (c) 2-cell $\sigma$ (red) and the set $\Sigma(\sigma)$ (blue) of possible intersection; (d) affine map of $\sigma\cup\Sigma$ on $z=0$ plane; 
   (e) reduction to a set of 1D segments in $\E^2$; (f) pairwise intersections; (g) regularized plane arrangement $\mathcal{A}(\sigma\cup\Sigma)$; (h) exploded 3-complex extracted from the 2-complex  $X_2(\cup_{\sigma\in B} \mathcal{A}(\sigma\cup\Sigma))$ in $\E^3$. The LAR representation of both the input data (not a complex) and the output data (3-complex with three 3-cells)  is given in Appendix~\ref{sec:appendix}
   }
   \label{fig:appendix}
\end{figure*}

\begin{figure}[htbp] 
   \centering
   \includegraphics[width=\linewidth]{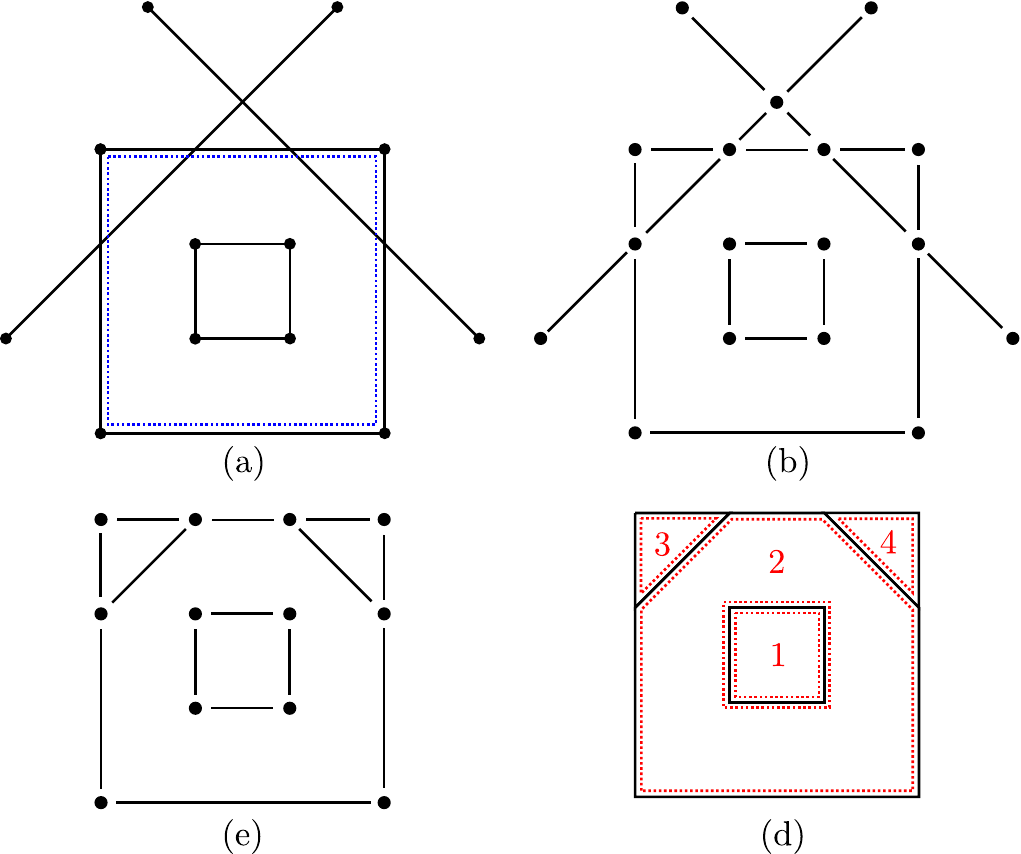} 
   \caption{Basic case: computation of the regularized arrangement of a set of lines in $\E^2$: (a) the input, i.e.~the 2-cell $\sigma$ (blue) and the line segment intersections of $\Sigma(\sigma)$ with $z=0$; (b) all pairwise intersections; (c) removal of the 1-subcomplex external to $\sigma$.
   To    this purpose an efficient and robust point classification algorithm~\cite{Paoluzzi-ART1986} is used,
   ; (d) interior cells of the regularized 2-complex $X_2=\mathcal{A}(\sigma\cup\Sigma)$  generated as arrangement of $\E^2$ produced by $\sigma\cup\Sigma$.}
   \label{fig:2D}
\end{figure}

\subsection{\texttt{LARLIB.jl} Design Goals}

We posit that basic data and algorithms used by the package may find a proper
fitting among the common architectures and representations of convolutional neural
networks, based on sparse matrices and linear algebra, in order to properly combine geometric modeling and image understanding.

In particular, our approach grounded on LAR to discover the
\(d\)-cells\footnote{Of very general type, even non convex and with any number of internal holes.}
of an unknown space arrangement~\cite{2017arXiv170400142P}, 
seems to match well with deep NNs~\cite{Goodfellow:2016:DL:3086952,Boxel:2016:DLT:3019358}. In
fact, our computation  proceeds by discovering in parallel the 0-cells, then by globally detecting the
incident edges, then by computing the equivalence relations and quotient
sets of 1-cells, then by proceeding to local (planar) reconstruction of
2-cells, and hence again to identification of quotient sets of 2-cells
in 3D, and finally to local (parallel) reconstruction of 3-cells, and so
on, even for higher dimensions.

\begin{figure*}[htbp] 
   \centering
   \includegraphics[width=\textwidth]{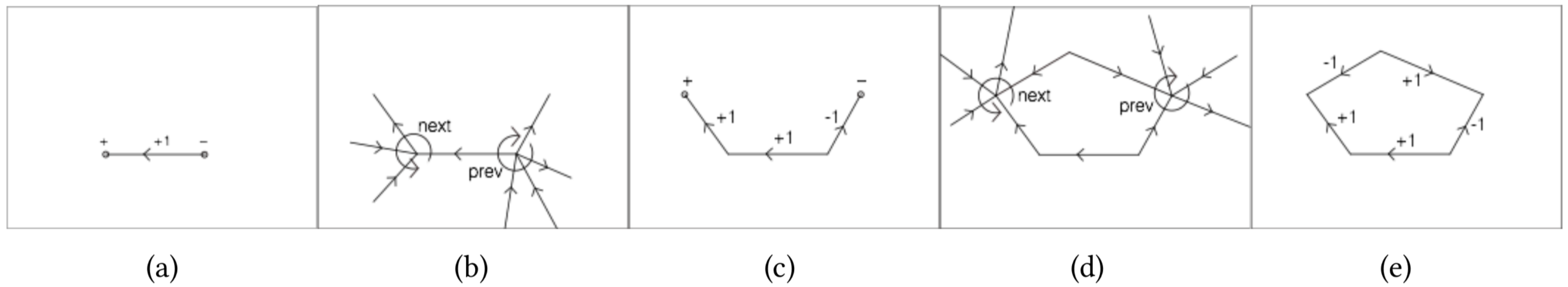} 
\caption{Dimension-independent topological gift-wrapping: extraction of a minimal 1-cycle from $\mathcal{A}(X_1)$: (a) the initial value for $c\in C_1$ and the signs of its oriented boundary; (b) cyclic subgroups on $\delta\partial c$; (c) new (coherently oriented) value of $c$ and $\partial c$; (d) cyclic subgroups on $\delta\partial c$; (e) final value of $c$, with $\partial c = \emptyset$.
 (Image from~\cite{2017arXiv170400142P})}
   \label{fig:gift2D}
\end{figure*}

\begin{figure*}[htbp] 
   \centering
   \includegraphics[width=\textwidth]{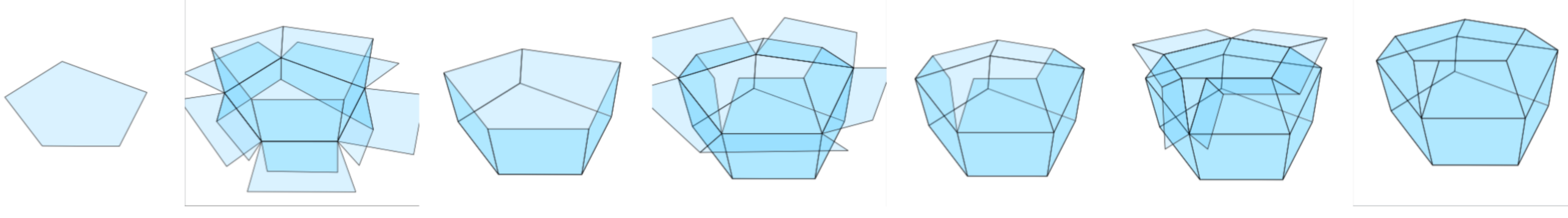} 
   
{\small~\hspace{0.04\textwidth}(a)\hfill(b)\hfill(c)\hfill(d)\hfill(e)\hfill(f)\hfill(g)\hspace{0.05\textwidth}.}
\caption{Dimension-independent topological gift-wrapping: extraction of a minimal 2-cycle from $\mathcal{A}(X_2)$: (a) 0-th value for $c\in C_2$; (b) cyclic subgroups on  $\delta\partial c$; (c)~1-st value of $c$; (d) cyclic subgroups on $\delta\partial c$; (e) 2-nd value of $c$; (f) cyclic subgroups on $\delta\partial c$; (g) 3-rd value of $c$, such that $\partial c=0$, hence stop. (Image from~\cite{2017arXiv170400142P})}
   \label{fig:gift}
\end{figure*}

When applied to very-high resolution imagery~\cite{doi:10.1080/16864360.2016.1168216} of biological structures
(e.g.~neuronal images~\cite{pao:cad16}) or to next-gen\-eration medical 3D imaging, this approach may produce 3D models of micro-level
structures, say capillary veins and small nerve structures, locally
reconstructing their dendritic meso-structures, including possible
inclusions of other components, and then proceed bottom-up to
hierarchically assemble more and more complex macro-structures.

\subsection{Package Core: (co)boundary matrices}

The main purpose of the \texttt{LARLIB.jl} core is the \emph{computation of the chain complex} induced by a cellular complex.  
In other words,  computing one/more sparse (co)boundary matrices, given the LAR representation of  one or more input cell complexes. Remember that a user-readable LAR representation is made by one (or two) characteristic matrices as array (indexed by cell numbers) of arrays of integer indices of incident vertices.  The reader may look at  \ref{sec:appendix} for a simple 3D example. 

When fed with the simplest user-readable representation of a characteristic matrix, i.e., with an array of arrays of positive integers, say with \texttt{FV} (faces-by-vertices) for a simplicial complex, augmented with \texttt{EV} (edges-by-vertices), in case of more general cells, and with the \texttt{V} array of vertex coordinates, the function \texttt{spaceArrangement()} returns either the sparse oriented matrix of the $\partial_3$ operator, or the basis \texttt{CV} of 3-cells of the space decomposition. Analogously, when the input provides either a 2D graph or a soup of 2D line segments, or both, and  is made by a \texttt{V} array of 2D vertex  coordinates,  and by the \texttt{EV} array of indices of vertices for each line segment (even not pairwise incident on vertices), the function \texttt{planeArrangement()} returns either the sparse oriented matrix of the $\partial_2$ operator, or the basis \texttt{FV} of 2-cells of the plane decomposition. 

\subsection{Package Core: the Merge Algorithm}

The computation of the arrangement of $\E^d$ 
generated by a collection of ($d$--1)-complexes, is called  \emph{Merge of complexes}.
This algorithm has a central role, since used for (a)  computing the (co)boundary matrices of a complex, (b)  testing/repair non perfect cellular complexes,  (c) mutually fragmenting a  collection of complexes before the reconstruction of Boolean expressions between solids.

We give here a simplified outline of it, since this algorithm is the actual core of the \texttt{Larlib.jl} implementation. For a more detailed exposition, the reader is referred to~\cite{2017arXiv170400142P}. Using dimension-independent language, we say that the input is a collection of cellular complexes of dimension $(d-1)$ embedded in the Euclidean space $\E^d$, while the output is the arrangement of $\E^d$ generated by them, represented as a single  cellular complex of dimension $d$. The current implementation is for $d\in\{2,3\}$.

\begin{description}

\item[Assemble the input complexes] 

First we assemble the $n$ input ($d$--1)-complexes in a single LAR representation $B$. This one is not a cellular complex, since cells may intersect outside of their boundaries. The purpose of the first part of the algorithm is to fragment the cells properly, so that they become a single ($d$--1)-complex embedded in $\E^d$.

\item[Compute a spatial index]

Then we efficiently compute a spatial index that associate to each ($d$--1)-cell $\sigma$ in the input, the set $\Sigma(\sigma)$ of ($d$--1)-cells with possible intersection with it. For this purpose we compute $d$ interval trees, allowing for fast computation of intersecting bounding boxes.

\item[Compute the facet arrangments in $\E^{d-1} $] 

\hfill
For each set $\Sigma(\sigma)$ of ($d$--1)-cells, embedded in $\E^d$, we compute the \emph{planar arrangement} generated by them in the closure (interior and boundary)  of $\sigma$. This arrangement is a ($d$--1)-complex having $\sigma$ as support space, i.e.~properly decomposing it in a cellular complex. 

\item[Assemble the output ($d$--1)-skeleton] 
\hfill
All planar arrangements $X_{d-1}^\sigma=\mathcal{A}(\sigma)$, for $\sigma\in B$, are syntactically merged into a single proper ($d$--1)-complex in $\E^d$, providing the ($d$--1)-skeleton of the output complex. This merging of $p$-cells ($0\leq p\leq d-2$) is performed via identification of quotient subcomplexes with a single instance of the common cells. The equivalence relation is computed looking for cells sharing the same portion of space, starting from identification of  vertices with the same coordinates (or very close, due to numerical round-offs). $Nd$-trees of vertices are used  for this purpose, and cells are written in canonical form (sorted arrays of unsigned vertex indices) to allow for syntactical identification.

\item[Extract d-cells from ($d$--1)-skeleton] Finally, the unknown $d$-cells of the output $d$-complex $X_d =\mathcal{A}(X_{d-1})$ are extracted. First note that a cycle is a chain with empty boundary. Each cell $\sigma_d$ is returned as a minimal ($d$--1)-cycle, extracted from $X_{d-1}$ using the \emph{Topological Gift Wrapping} algorithm displayed in Figure~\ref{fig:gift}. Each basis element of the chain space $C_d$, i.e.,~every $d$-cell, is represented as a ($d$--1)-cycle (see Section~\ref{sec:matrix}), and stored by column in the matrix $[\partial_d]$ of the linear operator $\partial_d: C_d\to C_{d-1}$.

    \begin{description}
    
    \item[Basis $d$-cells bounded by minimal ($d$--1)-cycles] The topological property used to compute the basis of $d$-cells is the fact that each ($d$--1)-cell is shared by at most two $d$-cells. The property holds exactly if considering also the exterior unbounded cell, that must be later discovered, with its column removed from the boundary matrix, since it is a linear combination of the other columns.
    
    \item[Topological gift wrapping] ~\cite{2017arXiv170400142P} reminds the gift wrapping algorithm~\cite{JARVIS197318}, but is more general, since it applies also to non-convex cells, and is multidimensional. It is implemented by iterating the application of the operator $\delta_{d-2}\circ\partial_{d-1}$, and starting from a singleton ($d$--1)-chain, as shown in Figures~\ref{fig:gift2D} and~\ref{fig:gift}.
    
    \item[Cyclic subgroups of the $\delta\partial c$ chain] \hfill The permutation of elements of the ($d$--1)-chain returned by the application of $\delta_{d-2} \circ \partial_{d-1}$ can be expressed as the composition of $m$ cyclic permutations, each one sharing one of ($d$--2)-elements of the boundary of the current chain value, where $m$ is the length of its boundary. In Figure~\ref{fig:gift} such \emph{corolle} can be subdivided into the sets of \emph{petals} around each \emph{hinge} boundary edge. 
    
    \end{description}

\item[Linear independence of the extracted {cycles}] \hfill The basis of elementary $d$-cycles corresponding to the \emph{interior} $d$-cells of the generated $d$-complex is linearly independent. The minimality of basis cycles is guaranteed by the fact that the cardinality of incidence relation between $d$-cells and ($d$--1)-cells is exactly twice the number of $(d-2)$-cells. For any other basis of cycles this cardinality is higher.

\item[Poset of isolated boundary cycles] 

When the boundary of the sum of all basis $d$-cells is an unconnected set of ($d$--1)-cycles, these $n$  \emph{isolated}  boundary components have to be compared with each other, to determine the possible relative containment and consequently their orientation. To    this purpose an efficient point classification algorithm is {used}, in order to compute the \emph{poset} (partially ordered set) {induced by the} containment relation of isolated components. 

\item[{Base case: arrangement of lines in 2D}] 

Most of the \\ actual numerical computing is performed in $\E^2$ by generating the 2D arrangement  generated by a set of lines $X_1^\sigma$ that, with $d=2$, provide exactly a collection of ($d$--1)-complexes. See Figure~\ref{fig:2D} for an illustration of the steps below. The computation described here is repeated for each $X_1^\sigma$ sub-complex generated from input data, i.e. for each 2-cell $\sigma$.

\begin{description}
    \item[Segment {subdivision}: linear graph] \hfill
    The set of line seg\-ments in $X_1^k$ is pairwise intersected. The endpoints and the intersection points become the vertices of a graph, and the segment fragments become the edges of it.
    
    \item[Maximal 2-point-connected subgraphs]  \hfill
    Using the Hopcroft-Tarjan algorithm~\cite{Hopcroft:1973:AEA:362248.362272} the maximal biconnected subgraphs are computed. The dangling edges and dangling trees are discarted.
    
    \item[Topological extraction of $X_2^\sigma$]  
    The topological gift wrapping algorithm in 2D  is used to compute the 1-chain representation of cells in $X_2^\sigma$, i.e.~its planar  arrangement, codified as $[\partial_2^\sigma]$. See Figures~\ref{fig:2D} and~\ref{fig:gift2D} for two graphical illustrations in 2D of the dimension-independent topological gift wrapping algorithm.
\end{description}

\end{description}

\begin{figure*}[htbp] 
   \centering
   \includegraphics[width=0.33\textwidth]{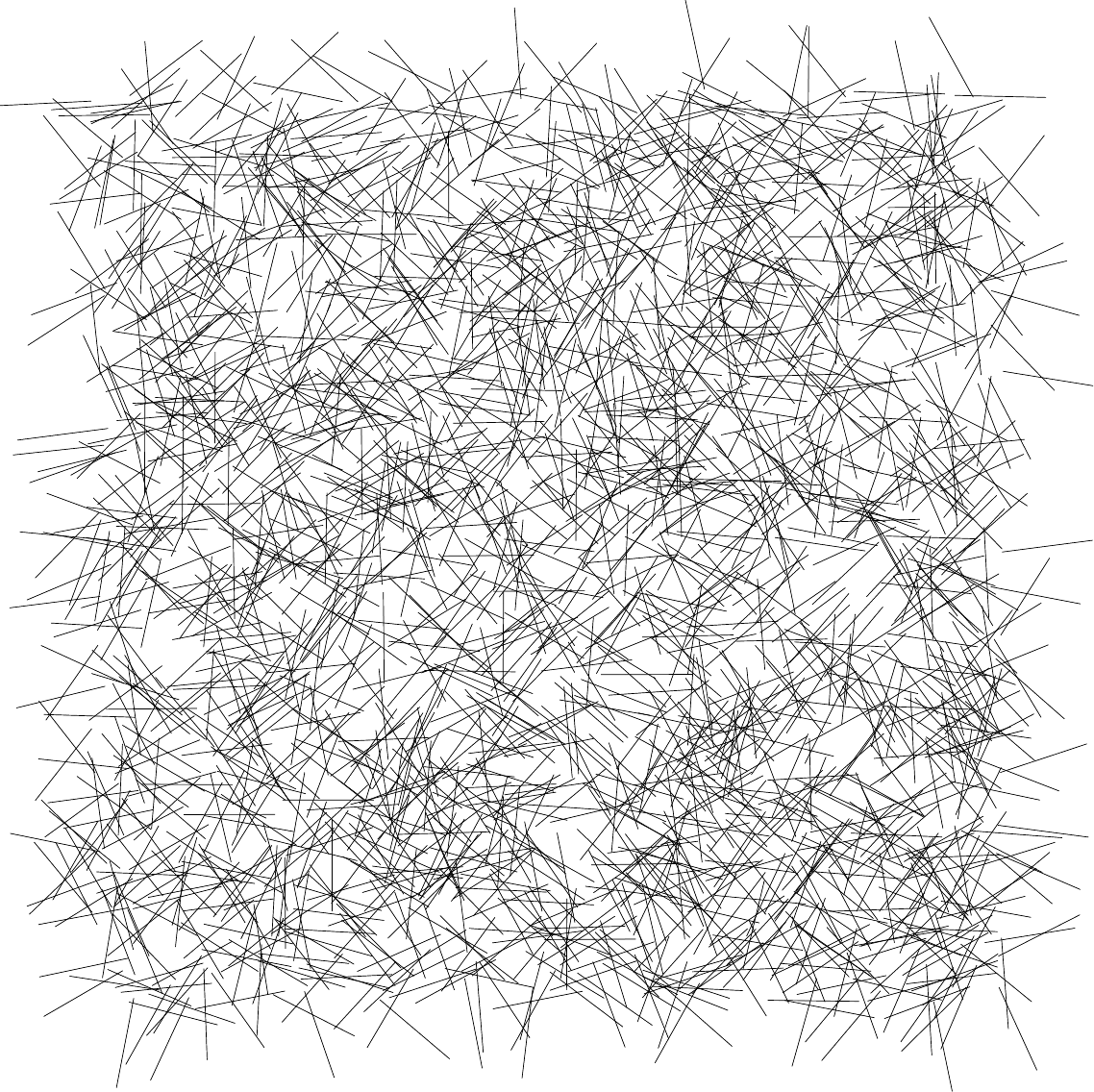}%
   \includegraphics[width=0.33\textwidth]{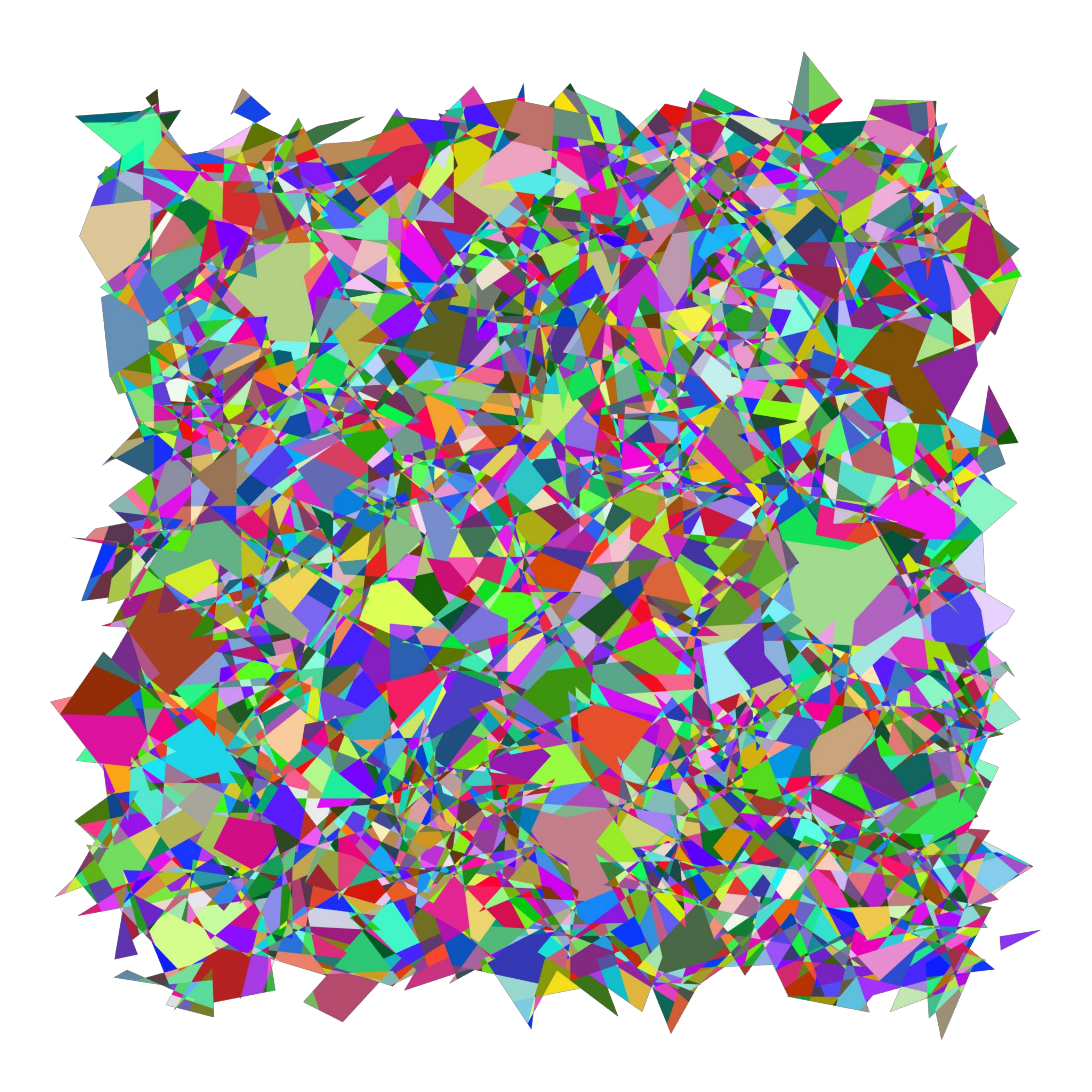}%
   \includegraphics[width=0.33\textwidth]{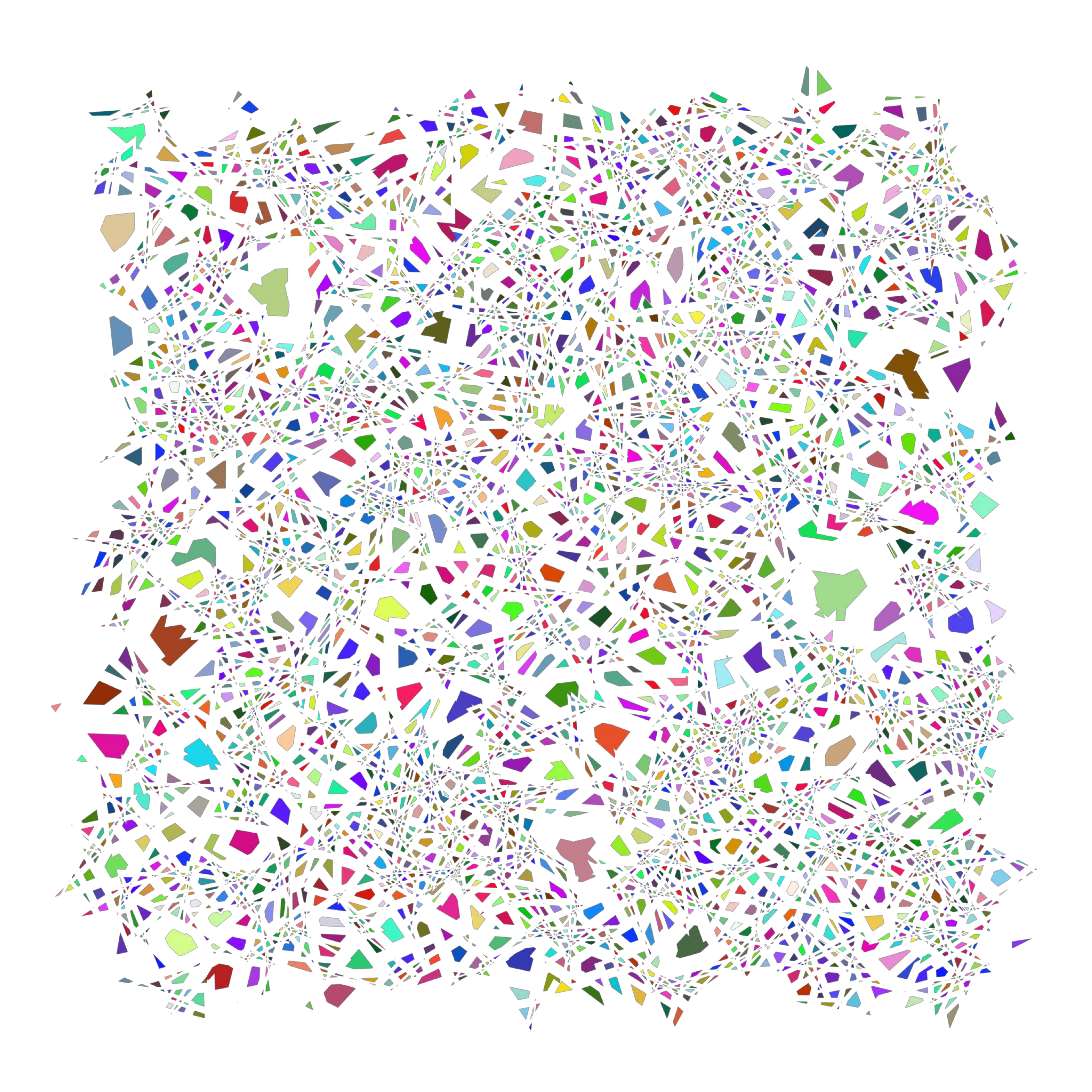}%
   \caption{The regularized 2D arrangement $X_2$ of the plane generated by a set of random line segments. Note that cells $\sigma \in X_2$ are not necessarily convex. The Euler characteristic is $\chi = \chi_0 - \chi_1 + \chi_2 =11361 - 20813 +  9454 = 2$.
   }
   \label{fig:example}
\end{figure*}

\section{Examples}\label{examples}
\subsection{Meaning of a Boundary Matrix}\label{sec:matrix}

{When constructing the matrix of a linear operator between linear spaces, say $\partial_d: C_d\to C_{d-1}$, by building it column-wise, we actually construct an element of the basis of the domain space, represented as a linear combination of basis elements of the codomain space. In this sense we ``build" or ``extract" one $d$-cell as a ($d$--1)-cycle, i.e.~by constructing its boundary or frontier. We would like to note that a \emph{cycle} ia a closed  chain, i.e.~a chain without boundary, and that the boundary of a boundary is empty.

The  \texttt{LAR} representation of $d$-cells, i.e.~the characteristic matrix $M_d$, is generated after the construction of the matrix $[\partial_d]$, by executing, for each column of $[\partial_d]$, and for each non-zero element of it (with signed indices of ($d$--1)-cells of the cycle), the  union of the  corresponding rows of the characteristic matrix $M_{d-1}$, so finally getting the list of vertices of the "built" or "extracted" $d$-cells. See the~\ref{sec:appendix} for the LAR representation of both the input (Figure~\ref{fig:appendix}a) and the output (Figure~\ref{fig:appendix}h) of the cartoon example in Figure~\ref{fig:appendix}, and for its coboundary  matrix $[\delta_2]=[\partial_3]^t$.}

\begin{figure*}[htbp] 
   \centering
   \includegraphics[width=0.3\textwidth]{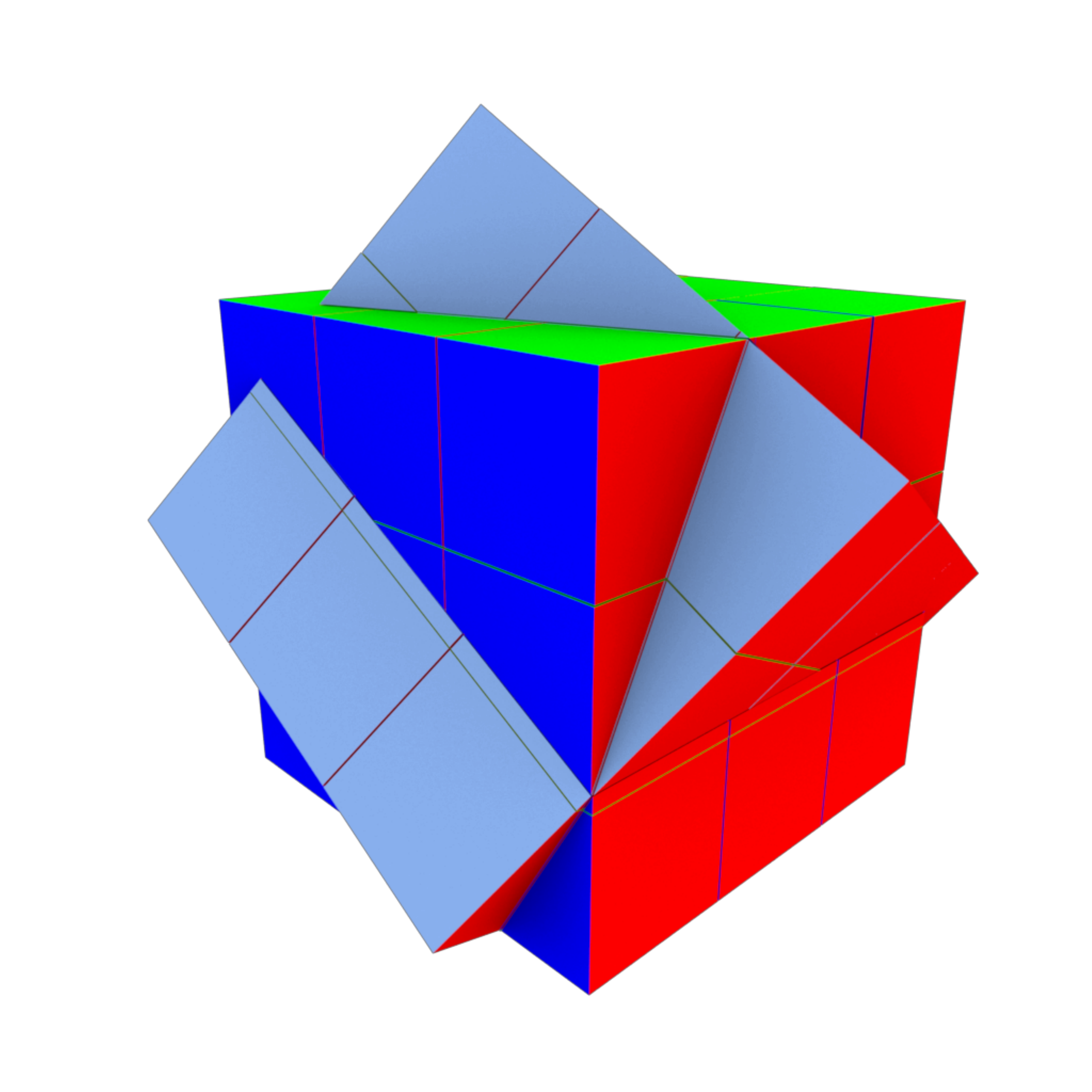}%
   \includegraphics[width=0.3\textwidth]{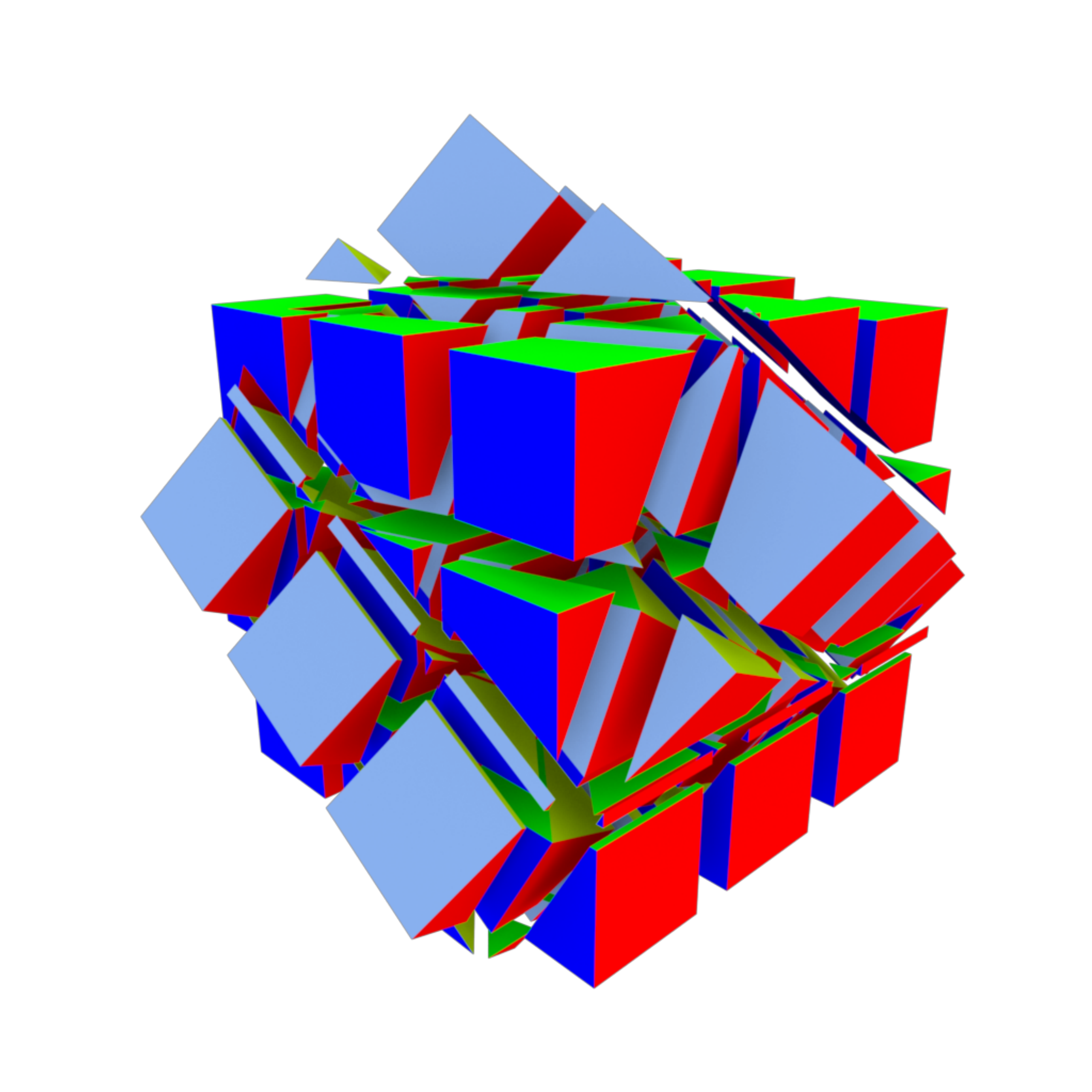}%
   \includegraphics[width=0.3\textwidth]{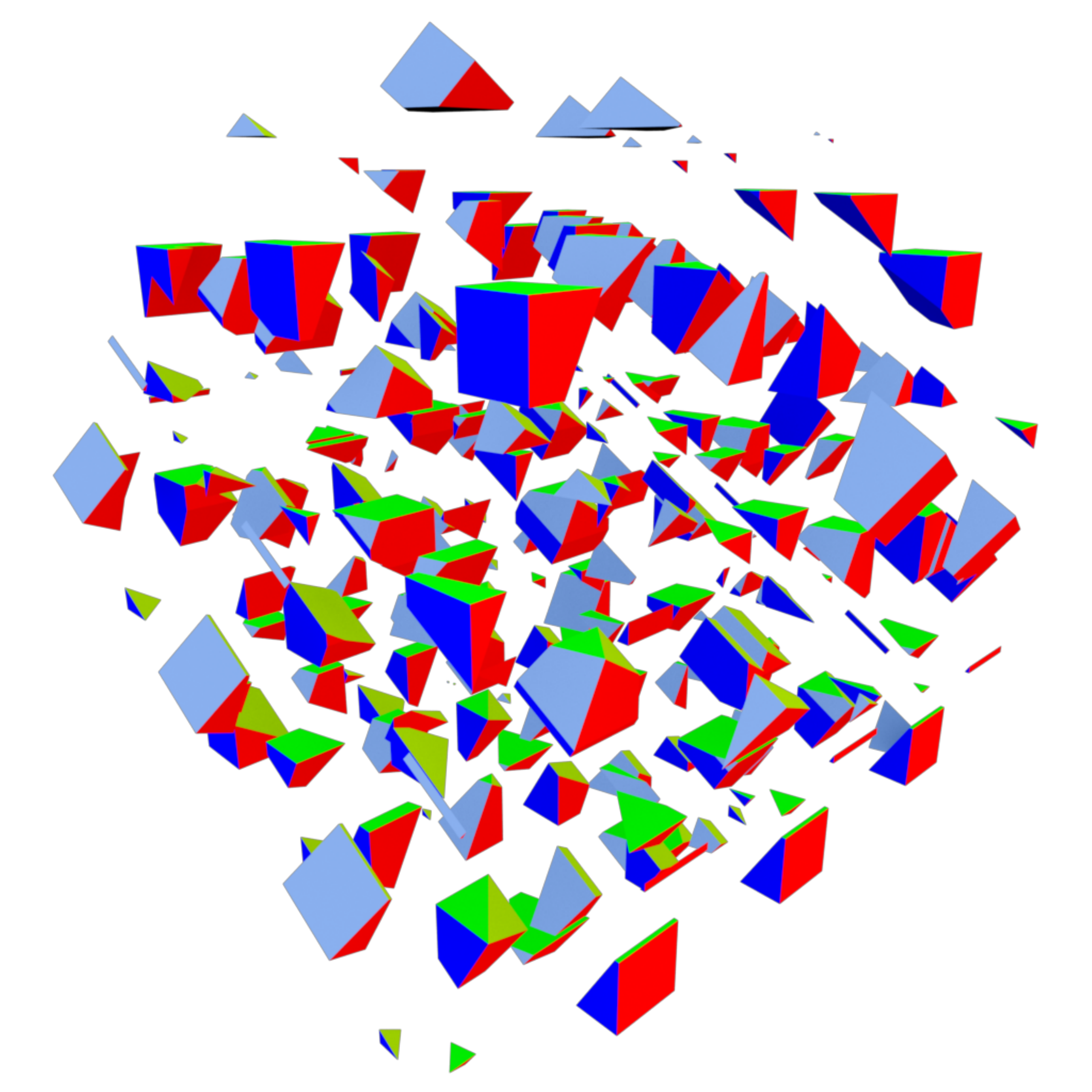}%
   \caption{Merge of two rotated $3\times 3\times 3$ cuboidal 3-complexes: (a)  the initial positions of the two input 3-complexes, each with $3^3$ unit cubic cells; (b)~3-cells of merged complex, with space explosion of small scaling  parameter; (c) larger space explosion. Each exploded cell is translated by a scaling-induced movement of its centroid. Output cells are not necessarily convex. In the merged complex we get 236 three-cells and 816 two-cells, both possibly non-convex. Ratio new/old 3-cells is $4.3704$.
   }
   \label{fig:rubik-1}

   \centering
   \includegraphics[width=0.3\textwidth]{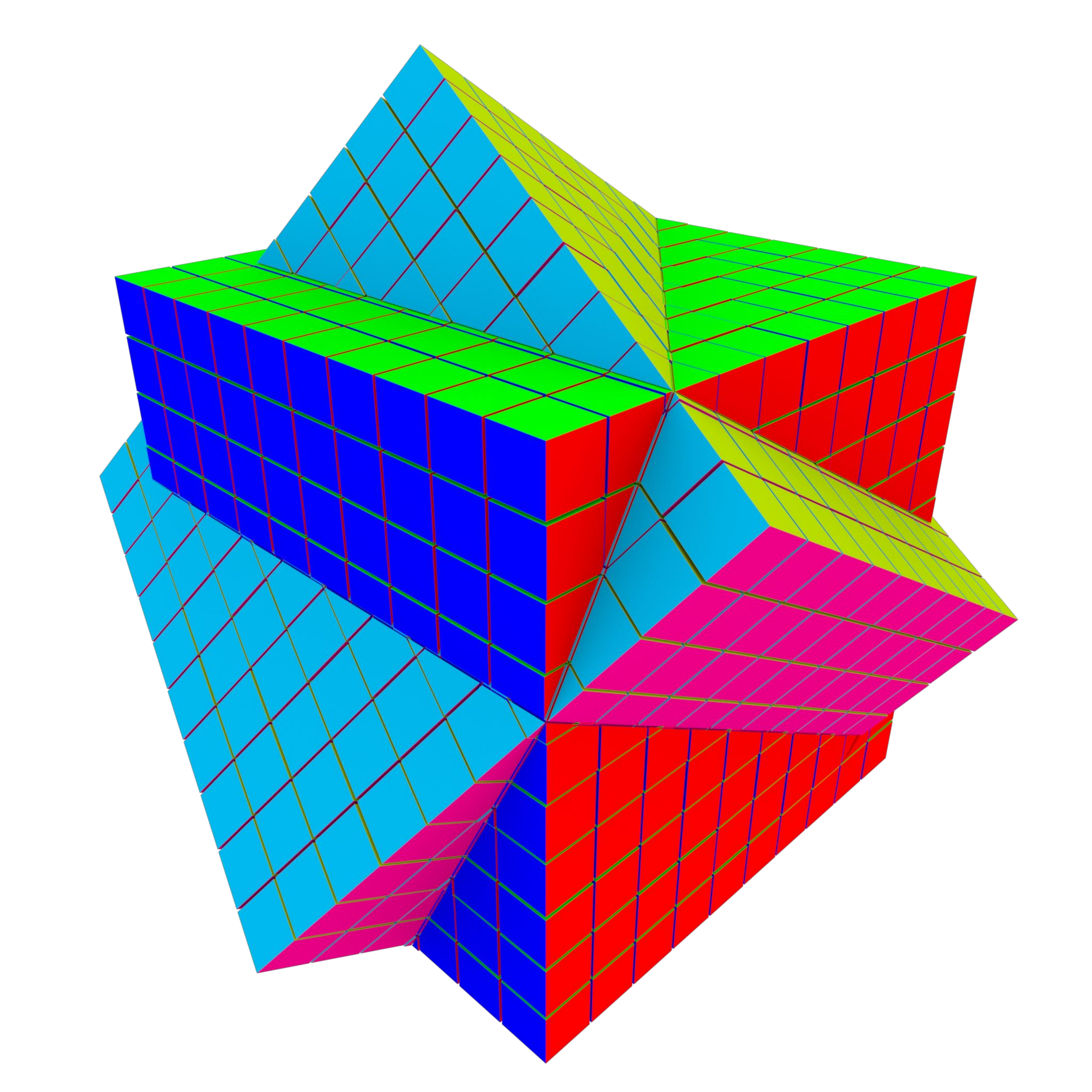}%
   \includegraphics[width=0.3\textwidth]{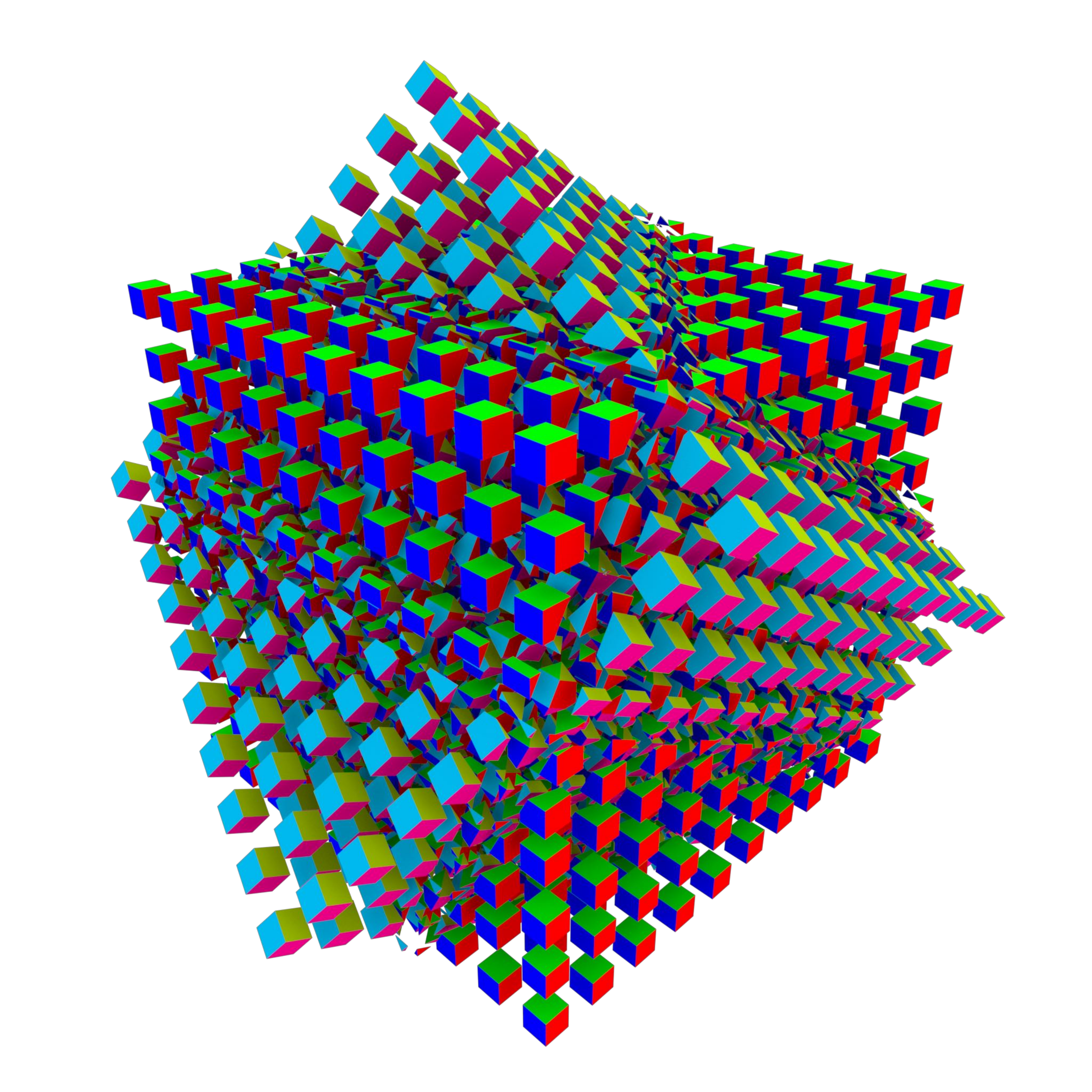}%
   \includegraphics[width=0.3\textwidth]{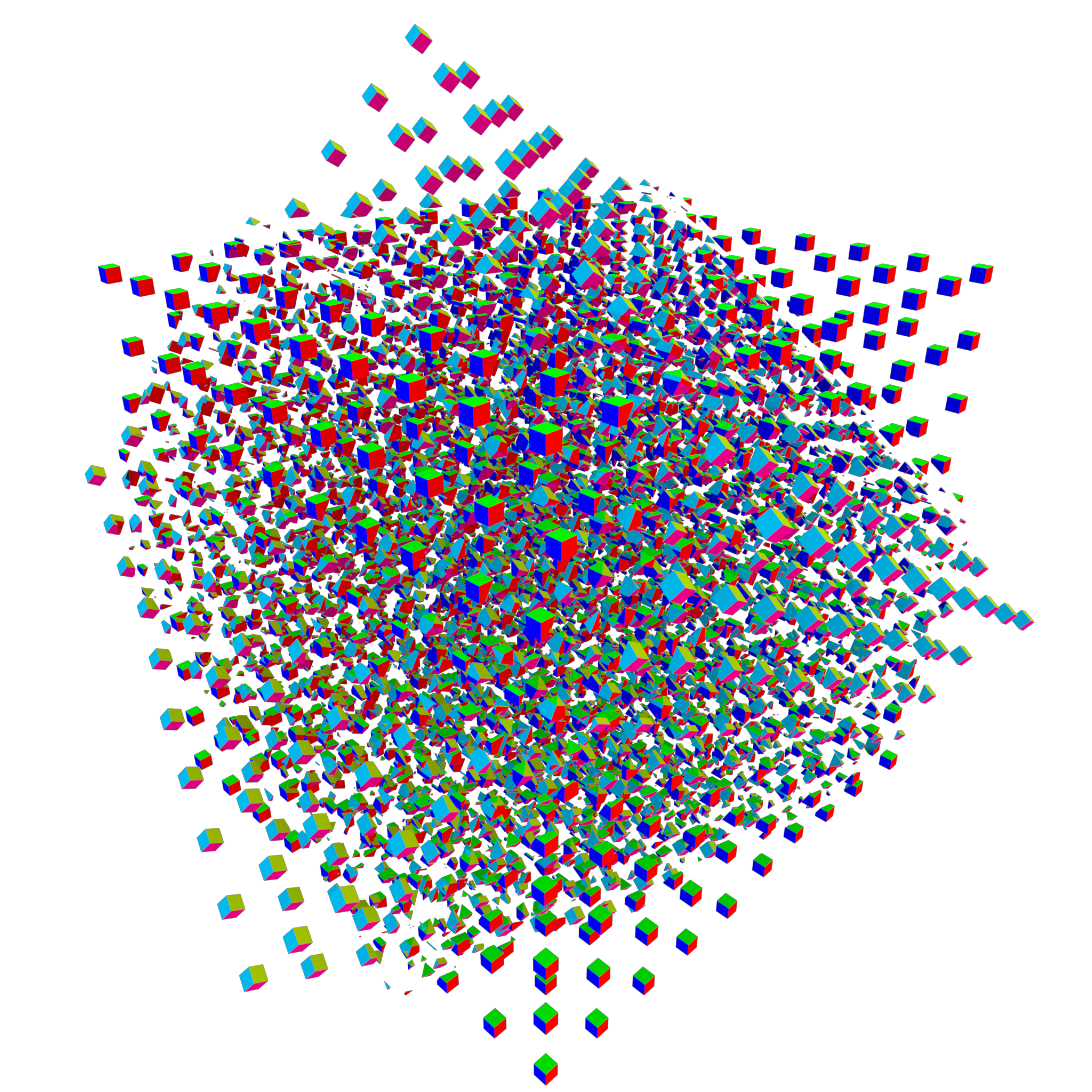}%
   \caption{Merge of two $10^3$  3-complexes. In the merged complex we get 8655 3-cells, 26600 2-cells (faces), 26732 1-cells (edges), and 8787 0-cells (vertices). Please note that the Euler characteristic is $\chi = \chi_0 - \chi_1 + \chi_2 - \chi_3 = 8787 - 26732 + 26600 - 8655 = 0$, since the decomposed 3-complex (no its exploded image!) is homeomorphic to the $n$-sphere, where  $\chi = 1+(-1)^n$, with $n=3$.  Ratio new/old 3-cells is $4.3275$.
   } 
   \label{fig:rubik-2}
\end{figure*}

\begin{figure*}[htbp] 
   \centering
   \includegraphics[width=\textwidth]{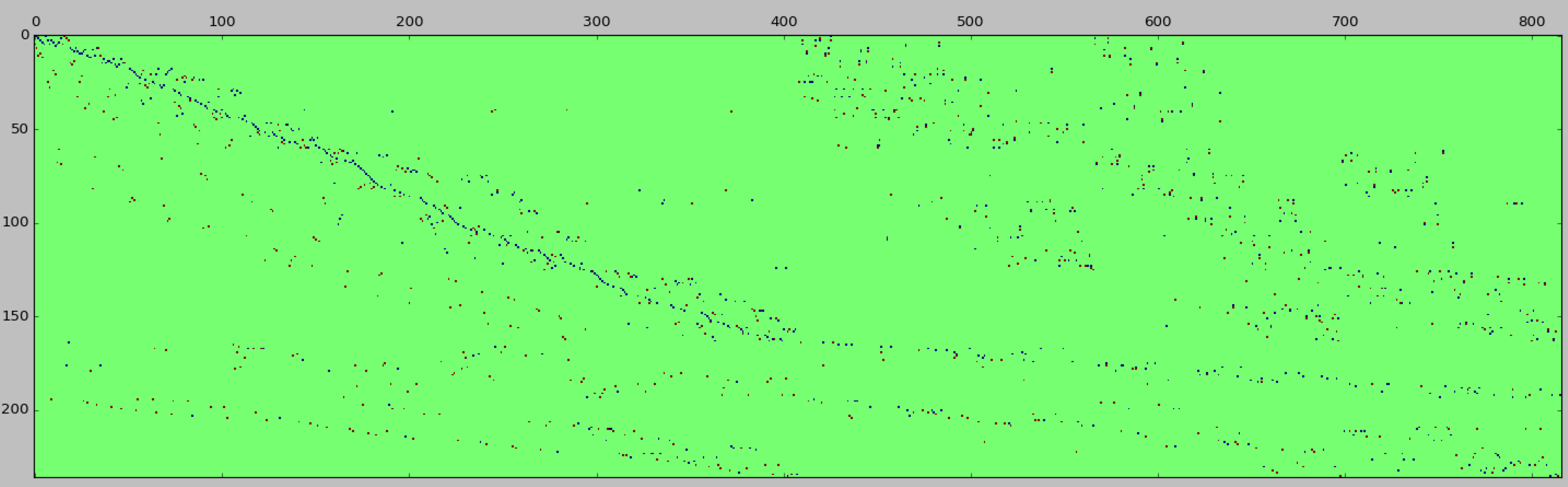}%
   \caption{Sparse matrix $[\delta_2]$ of the operator $\delta_2: C_2\to C_3$ for the cell-decomposition in Figure~\ref{fig:rubik-1}. We have a $(236\times 816)$ matrix of Julia's type \texttt{SparseMatrixCSC\{Int8,Int64\}} (CSC $\equiv$ Compressed Sparse Column) with 1.632 stored (non-zero) entries ${}\in\{-1,1\}$. Let us note that there are exactly \emph{two} opposite \emph{non-zero} elements in each column of $\delta_2$, external 2-cycle included, with each face providing a pair of non zeros. By enlarging a little portion to a close-up view, you can see unit values (blue) and their opposite (red) as small squares.
  }
   \label{fig:rubik}
\end{figure*}

\subsection{Merge of 3D cuboidal complexes}

A first example of the \emph{Merge} algorithm from the prototype implementation of \texttt{LARLIB.jl} is defined in this Section and displayed in Figure~\ref{fig:rubik-1}. The coboundary matrix $[\delta_2]$ is shown in Figure~\ref{fig:rubik}. 

The input data are two $3\times 3\times 3$ solid meshes of 3-cubes, both centered on the centroid, with the second rotated by  $\pi/6$ about the $x$ axis and by $\pi/6$ about the $z$ axis. Both the topologies are described in LAR as the pair \texttt{(FV,CV)}, i.e., as \emph{face-by-vertices} and \emph{cell-by-vertices}. As it is possible to check, there are $108\times 2$ quadrilateral faces and $27\times 2$ cubical cells in the input data. Of course, by column we have the indices to vertices of a single cell.

\begin{figure}
\scriptsize
\begin{verbatim}
julia> FV
4 x 108 Array{Int64,2}:
 1  2  3   5   6   7   9  10  11 ... 37  38  39  40  41  42  43  44
 2  3  4   6   7   8  10  11  12     41  42  43  44  45  46  47  48
 5  6  7   9  10  11  13  14  15     53  54  55  56  57  58  59  60
 6  7  8  10  11  12  14  15  16     57  58  59  60  61  62  63  64
\end{verbatim}
\caption{The 2-array description of the $M_2$ characteristice matrix of input data in Figure~\ref{fig:rubik-1}a, here with Julia type \texttt{Array\{Int64,2\}}, since all 2-cells are quadrilateral. The type becomes \texttt{Array\{\texttt{Array\{Int64,1\}},1\}} when the $d$-cells have a variable number of boundary vertices.}
\label{fig:FV}
\vspace{3mm}
\begin{verbatim}
julia> CV
8 x 27 Array{Int64,2}:
  1   2   3   5   6   7   9  10 ... 34  35  37  38  39  41  42  43
  2   3   4   6   7   8  10  11     35  36  38  39  40  42  43  44
  5   6   7   9  10  11  13  14     38  39  41  42  43  45  46  47
  6   7   8  10  11  12  14  15     39  40  42  43  44  46  47  48
 17  18  19  21  22  23  25  26     50  51  53  54  55  57  58  59
 18  19  20  22  23  24  26  27 ... 51  52  54  55  56  58  59  60
 21  22  23  25  26  27  29  30     54  55  57  58  59  61  62  63
 22  23  24  26  27  28  30  31     55  56  58  59  60  62  63  64
\end{verbatim}
\caption{The 2-array description of the $M_3$ characteristice matrix of input data in Figure~\ref{fig:rubik-1}a. Here all 3-cells are cubes, with 8 vertices for each.}
\label{fig:CV}
\end{figure}

The $p$-cells ($0\leq p\leq 3$) of the output 3-complex generated by the \emph{Merge} algorithm are actually codified inside the coboundary sparse matrices $[\delta_0], [\delta_1], [\delta_2]$. The algorithm actually codifies the (signed or unsigned, depending on the need) topological incidences called \texttt{(WE,EF,FC)}, where \texttt{W} is the array of vertices after the merge, i.e., the vertices of the fragmented cells, and their transposed matrices  \texttt{(EW,FE,CF)} respectively denote the \emph{Edges by vertices}, the \emph{Faces by edges}, and the \emph{Cells by faces}, in turn corresponding to $[\delta_0], [\delta_1], [\delta_2]$.

The sparse matrix $[\delta_2]$ of linear operator $\delta_2: C_2\to C_3$ is displayed as color image in Figure~\ref{fig:rubik}. Such $(236\times 816)$ matrix has Julia's type \texttt{SparseMatrixCSC\{Int8,Int64\}} and is stored in CSC (Compressed Sparse Column) format, with 1.632 stored (non-zero) entries in $\{-1,1\}$. 

It may be worthwhile to note that the  entries stored in $[\delta_2]$ are exactly the double of the 2-face number, i.e.~$2\,\chi_2$, since each 2-face provides in his column for exactly two (opposite) non-zero elements of $\delta_2$, external 2-cycle included. The same role would be played in a LAR-based Brep by the array $[\delta_1]=[\partial_2]^t$, with size $2\,\chi_1$.
Note also that the Brep of a solid would require at least size $8\,\chi_1$ (the main table of Winged-Edge Brep ).

\subsection{Posets of Unconnected Subcomplexes}

The \texttt{LARLIB.jl} package is providing a full implementation of the \emph{Merge} algorithm~\cite{2017arXiv170400142P}, including unconnected components in the output, either contained or not within some boundary shells of other components. To manage the very general case, the coboundary matrices of every isolated $d$-components are computed, each one including a redundant row (i.e.~linearly dependent on the other rows), and corresponding to the boundary ($d$--1)-cycle of the component. Such redundant $d$-cycles constitute a poset  with respect to the   relative containment relation. The poset's lattice is reconstructed via a point-shell containment algorithm, and the parity of each boundary $d$-cycle in the lattice graph is used to merge each container (even) and contained (odd) cycle pairs in the aggregated (co)boundary matrix .

\section{Conclusion}\label{conclusion}

In this paper we have discussed the design goals, the features and the implementation state of a  Julia package for topological computing with cellular and chain complexes. A distintive character of this approach is to deal with global operations over the space decomposition into connected cells, according to the old orientation of graphics hardware to give support to global/local affine or projective transformations. 
A $d$-chain is in fact a subset of cells of the same dimension $d$, and its coordinate representation is a sparse mapping ranging on the whole set of $d$-cells, and taking values either in $\{0,1\}$ or in $\{-1,0,+1\}$, depending on the possible consideration of cell orientation. Most topological operations and queries, and in particular the application of   boundary or coboundary operators, can be therefore performed on a subset (even the whole set) of cells at the same time, as a single SpMV (sparse matrix-vector) multiplication. This approach will favour the rewriting of many  topological/geometrical algorithms as dataflow processes of operator applications, according to the powerful hardware/software computational architectures of the last generation.

\subsection{How to Contribute}

The project discussed here is open-sourced and provides a liberal (MIT) licence, according to the common style of Julia's packages. New developers are needed in the short time to get the library to take momentum. The best approach is to fork the project repository on page
\texttt{https://github.com/cvdlab/LARLIB.jl} and use the current development release into your own project or application, then abstract some generally useful functionality, test it deeply, and finally ask for  pulling into the master branch. In case you use the software, please do not forget to mention this paper as a reference on your articles. 

\subsection{Next Steps}

Presently, the implementation reached a quite complete enactement of the 2D / 3D \emph{Merge} algorithm of cellular complexes~\cite{2017arXiv170400142P}, and generates a whole chain complex, i.e.~the whole set of (co)boundary operators acting on the space decomposition (i.e.~the space arrangement) induced by the merged set of input complexes. We are now working to its parallel optimization.
The next step concerns the use of core for reducing the resolution of any variadic Boolean formula between solids, including simple Boolean expressions, to the fast evaluation of a  predicate over a truth table.  The table contains by columns the coordinate representation of the input arguments within the space decomposition generated by the merge. The development plan also includes an efficient evaluation of the pseudoinverse of the boundary operator, in order to compute the field of distances from actual boundaries, to be applied over any fixed solid chain in the chain space\footnote{This operation is needed in order to perform mesh-free simulations of physical equations on cellular models.}. 
In addition, the students and collaborators of CVDLab at ``Roma Tre'' University started porting  to Julia  several modules of \texttt{Larlib.py}, including the applications related to extraction of geometric models from 3D medical images~\cite{doi:10.1080/16864360.2016.1168216}, and the design and/or material decomposition of building architectures~\cite{visigrapp17:cvdlab}.

\section*{Acknowledgements}
We gratefully acknowledge a partial support 2016-17 from SOGEI, the ICT company of the Italian Ministry of Economy and Finance.
A.P. would like to thank the friendly and warm assistance by Antonio DiCarlo and by Vadim Shapiro, that cooperated to establish our approach to discrete modeling, and to the foundations of LAR. 
Giorgio Scorzelli contributed by writing and  maintaining the \texttt{pyplasm} package, Enrico Marino and Federico Spini gave several types of support, also  by  testing some prototype ideas in Javascript.

\bibliographystyle{ACM-Reference-Format}

\appendix
\section{\uppercase{Appendix}}\label{sec:appendix}

\noindent The \texttt{Larlib.jl} code producing the example of Figures~\ref{fig:appendix}a and~\ref{fig:appendix}h, and the corresponding input and output arrays are given here.

{\footnotesize
\begin{verbatim}
V = [[0.0,0.0,0.0],[0.0,0.0,1.0],[0.0,1.0,0.0],[0.0,1.0,1.0],[1.0,
0.0,0.0],[1.0,0.0,1.0],[1.0,1.0,0.0],[1.0,1.0,1.0],[0.5,0.5,0.5],
[0.5,0.5,1.5],[0.0,1.366,0.5],[0.0,1.366,1.5],[1.366,1.0,0.5],
[1.366,1.0,1.5],[0.866,1.866,0.5],[0.866,1.866,1.5]]

FV = [[1,3,5,7],[9,10,11,12],[9,11,13,15],[2,4,6,8],[13,14,15,16],
[1,2,3,4],[3,4,7,8],[1,2,5,6],[9,10,13,14],[11,12,15,16],[10,12,
14,16],[5,6,7,8]]
\end{verbatim}}

The array \texttt{V} contains the input vertices; \texttt{FV} is the LAR representation of the characteristic matrix $M_2$, i.e. of the incidence relation between faces and vertices. The coboundary matrix $[\delta_2]: C_2\to C_3 = [\partial_3]^t$ between oriented chains, with $[\partial_3]: C_3\to C_2$, computed by the \emph{Merge} algorithm starting from the above data, is given below, with $[\delta_2]^t=[\partial_3]$:
\[
[\delta_2] = 
{\arraycolsep=2pt\def\arraystretch{0.8}
\mat{
 1 & 0 &-1 &-1 & 0 & 0 & 0 & 1 & 0 &-1 & 1 & 0 & 0 & 0 & 0 & 0 & 0 & 0\\
-1 & 0 & 0 & 0 & 0 & 1 &-1 &-1 & 0 & 0 &-1 & 1 & 1 & 0 & 0 & 1 & 1 & 0\\
 0 &-1 & 1 & 1 & 1 & 0 & 0 & 0 &-1 & 1 & 0 & 0 & 0 & 1 &-1 & 0 & 0 &-1\\
}}
\]

Since all rows of $[\delta_2]$ contain the representation of a 3-cell as a cycle of 2-cells, and analogously the $[\delta_1]$ and $[\delta_0]$ sparse matrices respectively contain the 2-chain basis (2-cells) as cycles of 1-cells, and the 1-chain basis (1-cells) as chains of 0-cells, it is easy to compute the LAR representation of the output complex, given below.

{\footnotesize
\begin{verbatim}
W = [[1.0,1.0,0.0],[1.0,0.0,0.0],[0.0,1.0,0.0],[0.0,0.0,0.0],[0.5,0.5,
1.0],[0.5,0.5,0.5],[0.2113,1.0,1.0],[0.2113,1.0,0.5],[0.5,0.5,1.5],
[0.0,1.366,0.5],[0.0,1.366,1.5],[1.0,0.7887,0.5],[1.0,1.0,0.5],[1.366,
1.0,0.5],[0.866,1.866,0.5],[1.0,1.0,1.0],[1.0,0.7887,1.0],[0.0,1.0,
1.0],[0.0,0.0,1.0],[1.0,0.0,1.0],[1.366,1.0,1.5],[0.866,1.866,1.5]]

FW = [[6,8,12,13],[9,11,21,22],[7,8,13,16],[12, 13,16,17],[10,11,15,
22],[3,4,18,19],[2,4,19,20],[5,6,12,17],[14,15,21,22],[5,7,16,17],[5,
6,7,8],[1,2,3,4],[5,7,17,18,19,20],[5,7,8,9,10,11],[8,10,12,13,14,15],
[1,3,7,8,13,18],[1,2,12,13,17,20],[5,9,12,14,17,21]]

CW = [[5,6,7,8,12,13,16,17],[1,2,3,4,5,6,7,8,12,13,17,18,19,20],[5,7,
8,9,10,11,12,13,14,15,16,17,21,22]]
\end{verbatim}}

It is worth noting the variable numbers of vertices per face in the \texttt{FW} array, with some non-convex faces, and the non-convexity of two solid cells in \texttt{CW} (see Figure~\ref{fig:appendix}h). The reader is kindly asked to compare for simplicity and compactness this representation with any other solid representation scheme. It also reduces to the standard mathematical representation of simplicial complexes, when cells are simplices. Anyway, LAR can be applied to more general cells, even containing holes.

\end{document}